\documentclass[nohyper,12pt,letterpaper]{JHEP3}
\usepackage[dvips]{epsfig}
\usepackage{amsfonts,amssymb}

\newcommand{\be}{\begin{equation}}
\newcommand{\ee}{\end{equation}}
\newcommand{\ben}{\begin{displaymath}}
\newcommand{\een}{\end{displaymath}}
\newcommand{\bea}{\begin{eqnarray}}
\newcommand{\eea}{\end{eqnarray}}
\newcommand{\bean}{\begin{eqnarray*}}
\newcommand{\eean}{\end{eqnarray*}}

\def\l {\lambda}
\def\a {\alpha}

\def\b {\beta}
\def\g {\gamma}

\def\d {\delta}
\def\s {\sigma}

\def\vp {\varphi}

\renewcommand{\t}{\theta}

\newcommand{\adss}[2]{\mbox{$AdS_{#1}\times {S}^{#2}$}}

\newcommand{\bra}[1]{\mbox{$\langle #1 |$}}
\newcommand{\ket}[1]{\mbox{$| #1 \rangle$}}

\newcommand{\eg}{{\it e.g.}}
\newcommand{\ie}{{\it i.e.}}

\newcommand{\tr}{\mbox{Tr}}

\newcommand{\commentout}[1]{}

\newcommand{\beq}{\begin{equation}}
\newcommand{\eeq}{\end{equation}}
\newcommand{\beqr}{\begin{displaymath}}
\newcommand{\eeqr}{\end{displaymath}}
\newcommand{\beqa}{\begin{eqnarray}}
\newcommand{\eeqa}{\end{eqnarray}}
\newcommand{\beqar}{\begin{eqnarray*}}
\newcommand{\eeqar}{\end{eqnarray*}}
\renewcommand{\k}{\kappa}

\newcommand{\p}{\phi}
\newcommand{\cN}{{\cal N}}

\newcommand{\cO}{{\cal O}}

\newcommand{\non}{\nonumber}

\newcommand{\half}{\ensuremath{\frac{1}{2}}}

\newcommand{\bi}{\ensuremath{\bar{\imath}}}

\newcommand{\bl}{\ensuremath{\bar{l}}}
\newcommand{\ba}{\ensuremath{\bar{a}}}

\newcommand{\N}[1]{\ensuremath{\cN=#1}}

\newcommand{\tl}{\ensuremath{\tilde{\lambda}}}

\newcommand{\sql}{\ensuremath{\sqrt{\lambda}}}
\newcommand{\bv}{\ensuremath{\bar{v}}}

\newcommand{\rf}[1]{(\ref{#1})}

\def \ci{\cite}

\def \const {{|rm const}}

\def\be{\begin{equation}}
\def\ee{\end{equation}}
\def\ba{\begin{eqnarray}}
\def\ea{\end{eqnarray}}

\def\a{\alpha}
\def\b{\beta}

\def\p{\partial}

\def\tr{{\rm  tr}}
\def \del{\partial}
\def \a {\alpha}

\def\g{\gamma}
\def\s{\sigma}

\def\ov{\over}
\def\la{\label}

\def\J{{\cal J}}

\def\LL{{\cal J }}

\def\E{{\cal E}}

\def\b{\beta}
\def\l{\lambda}

\def \adss{$AdS_5 \times S^5$\ }

\def \t {\theta}
\def \vp {\varphi}

\def \ov {\over}
\def \s{\sigma}

\def \ta{\tau}

\def \ha {{1 \over 2}}

\def \la{\label}

\def \k {\kappa}
\def\foot{\footnote}

\def\det{\hbox{det}}
\def \ci {\cite}

\def \foot {\footnote}
\def \bi{\bibitem}
\def \tr {{\rm tr}}
\def \ha {{1 \over 2}}
\def \tid {\tilde} 

\def \tl {{\tilde \l}}

\def \ta {{\tilde \a}}
\def \fo { {1\ov 4}}
\def \ep {\epsilon}
\def \inti {{\int^{2\pi}_0 {d \sigma \ov 2 \pi}}}
\def \be {\bea}
\def \ee {\eea}
\def \d {\partial}

\def \l  {\lambda}
\def \tl {{\tilde \l}}
\def \bl {{\tilde \l}}
\def \const {{\rm const}}
\def \V {v}

\def \bv {v^*} 
\def \vv {{\rm v}}
\def \LL {{\cal L}}
\newcommand{\PV}[1]{P_{\!\!_{V_{#1}}}}
 
\def \bL {\ell} 
\def \galpha {\gamma} 
\def \M {{\cal M}}

\title{\LARGE 
Semiclassical relativistic  strings in $S^5$ 
and long coherent  operators in
$\cal N$=4 SYM  theory
}

\author{
Martin Kruczenski  \thanks{E-mail:martink@brandeis.edu} \\
        Department of Physics, Brandeis University \\
       Waltham, MA 02454, USA.
}

\author{
A.A. Tseytlin \thanks{Also at Imperial College London
and  Lebedev   Institute, Moscow.} \\
 Department of Physics,\\
The Ohio State University\\
Columbus, OH 43210-1106, USA.
}

\abstract{
We consider the low energy 
effective action corresponding to the 1-loop, planar, dilatation operator
 in the scalar sector of $\N{4}$  $SU(N)$ SYM theory.
For a general class of non-holomorphic ``long'' 
operators, of bare dimension $L\gg1$, it is a 
sigma model action with  8-dimensional 
target space and agrees with a limit of the phase-space string sigma model 
action describing generic fast-moving  strings in 
the $S^5$ part  of $AdS_5 \times S^5$. 
The limit of the string action is taken in 
 a way that allows for a systematic expansion to higher orders
in the effective coupling  $\tilde \lambda = {\lambda \ov L^2}$. 
This extends  previous work on rigid  rotating strings in $S^5$
(dual to operators in the $SU(3)$ sector of the dilatation operator)
to the case when string oscillations or pulsations 
in $S^5$ are allowed. 
We  establish a  map between
 the profile  of the leading order string solution and 
the structure of the corresponding coherent, 
``locally BPS'', SYM scalar operator.

As an application, we explicitly determine the form of  the 
 non-holomorphic operators dual  to the pulsating strings.
Using   action--angle variables, we also directly  compute
 the energy of pulsating solutions,   simplifying 
previous treatments.

}

\preprint{\tt{BRX TH-543} \\
          \tt{hep-th/0406189}  }

\begin{document}

\section{Introduction}

The AdS/CFT correspondence \cite{malda} gave
 a precise example of the conjectured
relation \cite{largeN} between the large $N$ limit of
 gauge theories and string theory.
In its most well-known form it claims  that 
\N{4} SYM theory with gauge group $SU(N)$ 
and coupling $g_{_{\rm YM}}$ on one hand, and type IIB string
 theory on \adss with $N$ units of RR 5-form flux,  
and string coupling $g_s=g_{_{\rm YM}}^2$ on the other, 
are just two different descriptions of the same theory.  
The  string theory becomes weakly coupled, i.e.
 the 
theory becomes ``stringy'',  
in the limit $g_{_{\rm YM}}\rightarrow 0$, 
$N\rightarrow \infty$,  with  the 't Hooft 
coupling $\lambda=g_{_{\rm YM}}^2N$ being
  fixed and large,   $\lambda\gg 1$.

It remained unclear, however,   how strings
 ``emerge'' from the field theory,
 in particular,  which (local,  single-trace) 
 gauge theory operators \ci{POL} should 
 correspond to which ``excited'' string states and 
 how one may   verify
the matching of  their
 dimensions/energies 
beyond the well-understood BPS/supergravity sector.  
An important step in that direction was made 
in 
\ci{bmn} where it was shown  how this  correspondence can be 
established for  a class of ``small'' (nearly point-like)  near-BPS
strings  which are ultrarelativistic, i.e.  whose  kinetic energy 
is much larger  than  their mass.

Shortly after,  in \cite{gkp} it  was argued that at least a qualitative
 agreement between the non-BPS states on the two sides of the duality 
can be established  also 
for certain  extended string states represented by 
classical string solutions with one large $AdS_5$ 
angular momentum.
The semiclassical approach of \ci{gkp} was
further developed and generalized to 
multispin string states in \cite{ft1,ft2}. It 
 was proposed  in \ci{ft2} that,  like for ``small'' near-BPS  
BMN \ci{bmn} string states, 
a {\it quantitative} agreement 
between string theory and gauge theory 
should be found also for a class of 
extended classical solutions with two or three 
non-zero  angular momenta in the $S^5$ (see \ci{tse2}
for a review). For such solutions the classical energy has a regular 
expansion in powers of $\tl = {\l \ov L^2}$,  where $L=J$ is the total 
$S^5$ spin, $E=L(1 + c_1 \tl + c_2 \tl^2+ ..)$, and the   
string $\a'\sim {1\ov \sql}$ 
corrections are suppressed in the limit $L \to \infty, 
\ \tl=$fixed. 
Assuming that the large $L$ and then small  $\tl$ 
limit is well-defined also on the SYM side, one should 
then be able 
to compare the classical string results for the energy 
 to the  quantum SYM results for the  corresponding anomalous
 dimensions.

Using  the crucial observation of \ci{mz1}
that the one-loop scalar dilatation operator 
can be interpreted as a Hamiltonian of an integrable $SO(6)$
 spin chain and thus can be diagonalized even for large $L$ 
by the  Bethe ansatz method, the 
proposal of  \cite{ft2} was confirmed at the leading order 
of expansion in $\tl$ in \cite{bmsz,bfst}. There, a 
 remarkable agreement 
was found between energies of  various string solutions and 
eigenvalues  of the dilatation operator representing
 dimensions of particular SYM operators.



The established correspondence  was  thus   between a 
 thermodynamic limit  $L \to \infty$ of the 
Bethe ansatz eigenstates of the  integrable spin chain 
and a large spin, or, equivalently, 
 large energy,  limit of the classical solitonic solutions 
of the  string sigma model action. 
The classical bosonic coset  sigma model  also  has a
well-known  integrable 
structure which  becomes  explicit 
for particular rigid-shape rotating string configurations
\cite{afrt,art} and which can be mapped \cite{as,Min2}
to the one of the spin chain.

The next important step was made  in  \cite{kru} where it was shown  
that one can take the large energy, or ultrarelativistic, limit directly 
in the string action getting  a reduced ``non-relativistic'' 
sigma model that describes in a universal way 
the  leading-order $O(\tl)$ corrections 
to the energies of all string solutions in the two-spin sector.
Moreover, it was found  \ci{kru} that the resulting action 
agrees  exactly with   the semiclassical 
 coherent state action describing the 
$SU(2)$ (Heisenberg XXX$_{1/2}$) sector of the  spin chain
in the $L \to \infty, \ \tl=$fixed  limit, which turns out to be 
equivalent to a
continuum limit in which one keeps only linear and quadratic 
derivative terms.  
This demonstrated  how a string action can directly emerge from a gauge 
theory in the large-$N$ limit and 
provided a direct map between  
``coherent''  SYM  states or operators built out of two holomorphic  
scalars, and all two-spin classical string states,
 bypassing the need 
to apply the Bethe ansatz to find anomalous dimension
for each particular state. 
Furthermore, the correspondence  established at the level
of the action implies also  the matching of fluctuations 
around particular  solutions
 (just as in the BMN case where one matches 
fluctuations near a BPS state) 
and thus  goes 
 beyond the case of  rigidly rotating strings. 
This remark  applies also to other subsectors which we shall 
 describe below which
therefore overlap, i.e. are  related by small deformations. 
These subsectors  are named according  to a certain type of basic 
solutions they contain (which carry several large conserved charges)
but they also describe many  ``nearby'' 
states which may be labeled by higher conserved charges 
or oscillation numbers.


The observation made  in \ci{kru} may 
be  viewed  also  as based on the fact 
that the spin chain has two equivalent descriptions,
 a   Hamiltonian operator one and a 
Lagrangian  path integral one. 
The corresponding   Lagrangian  was shown to be  identical to a
 limit of the classical string Lagrangian. 
The relevant semiclassical limit of the path integral 
is, as usual,   naturally represented   by  the coherent states
of the operator approach. 
The matching demonstrated  in \ci{bmsz,bfst} can then be interpreted 
as an equivalence  between  the two descriptions of the spin chain. 
The remarkable fact is 
 that the classical solutions in 
 the path integral approach, namely 
the solutions of the Landau-Lifshitz (LL) equations, 
are in correspondence with  exact eigenstates 
found using the thermodynamic limit of the  Bethe ansatz: the 
energies as well as  all higher conserved charges  are the same. 
 The general proof of this fact  was given 
later in \ci{kmmz} using integrable models methods.

 From the effective action point of view,  one
 can also argue that, to  lowest order
in derivatives, there is only one unknown coefficient that can be fixed, e.g., 
 by
 comparison
with the BMN \ci{bmn} result.
 Therefore, at leading order in the $L \to \infty,
 \ \tl=$fixed  limit,
the effective action is unique and should be expected to reproduce the 
same limit of the exact results.
This uniqueness is lost at higher orders in $\tl$ where more terms 
in the effective action are present.
In that case, the coherent state approach needs to be generalized  as 
was explained  in \ci{krt}.
This allowed us to verify the correspondence at the $\tl^2$ 
order. An equivalent general  result 
(using the integrable spin chain embedding of \ci{SS})
was obtained  also  in  the Bethe ansatz 
approach \ci{kmmz}.

The approach of \ci{kru,krt} 
 was also generalized (to leading order in  $\tl$) 
to the three-spin or $SU(3)$ sector \cite{lopez, ST} 
(as well as to the $SL(2)$ \ci{BS,bfst} sector 
corresponding to one spin in $AdS_5$ and one in the $S^5$ \ci{ST}).
On the  Bethe ansatz side the agreement between the energies of particular 
3-spin string 
solutions    and the corresponding spin chain eigenvalues 
was previously shown  in \ci{Min2,char}.


The results reviewed  above 
explained the matching between all $S^5$ rotating
string  solutions with 3 large angular momenta 
 and all ``long'' operators constructed out of the three 
scalars $X=X_1+iX_2$, $Y=X_3+iX_4$,  $Z=X_5+iX_6$
(including, as mentioned above, also  ``near-by'' states).
However, there is also 
another interesting class of solutions for strings moving in $S^5$ 
--   the so called 
pulsating strings  \ci{Min1,Min2} which also have a regular
 expansion of their energy in terms $\tl = {\l \ov L^2}$ 
where $L$ is  a large ``oscillating number''.   Their energies 
 were matched (to order $\tl$) to the energies of the SYM theory states 
in \ci{bmsz,Min2} 
using the  Bethe ansatz techniques for the $SO(6)$ spin chain 
of \ci{mz1}. 


To carry out a 
similar matching  at the level of the effective action,
 that is to match  the corresponding 
coherent  states and not only the energy eigenvalues,\foot{The  
Bethe ansatz approach \ci{mz1,bmsz,Min2} provides, in principle, 
a recipe to construct the corresponding pure eigenstates or Bethe wave functions, 
but this is not easy to do in practice.}
 we need to understand how to 
extend  the
 ideas discussed above in the rotating string $SU(3)$ sector 
 to the whole $SO(6)$ spin chain \ci{mz1}, i.e.  to 
 the subset  of the SYM operators 
constructed out of all 6 real scalars and  
not limited to the holomorphic 
products of  $X$, $Y$, $Z$.

 The first step towards that goal was made 
 in \cite{ST}. There, the Grassmanian 
$G_{2,6}=SO(6)/[SO(4)\times SO(2)]$ was identified 
as the coherent state 
target space for the spin chain sigma model since it 
parametrizes the orbits of the  half-BPS operator
$X=X_1+iX_2$ under the $SO(6)$ rotations. 
In a related development, motivated by the suggestions in 
 \ci{mateos} and 
\ci{kru},  the procedure of taking the high energy or $\tl \to 0$ limit 
of the classical string theory  was generalized 
 \ci{mik,mikk} to the whole \adss bosonic action 
(and was later  extended to include fermions \ci{mikkk}).

\bigskip\bigskip 

 In the present paper we shall 
study the  $SO(6)$  sector in detail,  carefully 
working out  the spin chain and the  string theory sides 
of the correspondence. 
We will  
 show that the agreement between 
the two effective 
actions extends to the whole subsector of scalar 
operators characterized by a ``local BPS'' condition,  
i.e. built out of products of 
$SO(6)$ rotations 
of the  BPS  6-vector $(1,i,0,0,0,0)$. It is this 
condition  that selects the coset $G_{2,6}$ 
as the target space. 
This condition 
ensures 
that the corresponding  anomalous dimensions 
on the field theory side are  of order
$\lambda\ov L$ and thus 
can be compared to  the leading order corrections to the 
energy  on the string side. 
The role of this  locally  BPS  condition was also emphasized  in 
\cite{mikk}.

 On  the string side,  we need to find a  ``reduced'' 
sigma model by taking a  large energy
limit of the classical  string action. 
We  shall essentially follow \cite{kru,krt,ST} but 
 improve the derivation of
the reduced action in  two ways. 
 First,  we will 
clarify the gauge fixing procedure by using an alternative, 
2-d  dual (or ``T-dual'') action where
 the linear in time derivative ``Wess-Zumino''
 term appears more naturally from the usual 
$B_{mn}$--field coupling term. 
Second,  we will use 
canonical transformations to systematize the change of variables
 that was previously needed \ci{krt}
to eliminate 
terms of higher than first power of  time derivatives.
 In this way will 
we find a completely systematic and universal 
procedure to derive higher order in $\tl$ 
 corrections on the string theory 
side.
 The procedure
is  independent of a  particular solution one may 
 consider and, 
moreover, 
it should be possible to generalize this   procedure 
to the full \adss case.

We should mention  that the method of canonical 
perturbations was already applied to this problem 
in  ref. \cite{mikk} which  computed the leading 
order  action for a generic fast motion in \adss 
using a similar but somewhat 
 different approach  based on consideration of 
near 
light-like surfaces. The advantage of our procedure 
 is in its systematic nature which 
allows us to compute higher order corrections
with relative ease.

On the spin chain side,  we will use a variation of the coherent state 
approach that was  successful in the $SU(2)$ and $SU(3)$ case.
In the coherent state approach, 
one first reformulates the quantum mechanical 
 spin chain problem in terms of a coherent state path integral  \cite{fra} 
and then observes that in  the  limit we are interested in, i.e.
$L\to \infty, \ \tl=$fixed,  one can take the continuum limit
and all quantum corrections can be ignored. 
 This is essentially equivalent to ignoring quantum mechanical
correlations between different sites of the spin chain,  and,
as a result, we are lead to a classical action for the system.
 The variation of the coherent state approach 
we shall use is simply to ignore quantum correlations from
the very beginning by considering states
 which are the product of independent 
states at each site. The classical action 
can then be thought of as  
the action that leads to the Heisenberg equations of motion 
in this restricted subspace. 
This method, which is equivalent to  the one used
in \ci{krt}, leads to the same result at leading order but has some 
practical advantages when applied at  the next order.


In this paper we will only consider the leading 
order term in the spin chain effective action 
 which turns out to be in perfect 
agreement with the leading order action obtained on the string side.


An important consistency  check 
is  that, starting  with the reduced sigma model, 
we should be  able to reproduce 
the leading order results for all pulsating and
 rotating string solutions in  $R_t \times S^5$. 
The map that emerges 
between the field theory and the string theory \cite{kru,mik,mikk} 
is that each portion of the string that moves approximately
along a maximum circle carrying one unit of R-charge, corresponds 
to a site of the spin chain at which 
 there is a half-BPS operator with the same R-charge. 
 In particular, this map allows us to find the coherent 
operators corresponding to the pulsating strings of \cite{Min1,Min2}.
 Since the pulsating solutions are 
a particular case, the agreement between the 
actions that we find here 
explains the agreement already observed in 
\ci{bmsz,Min2} between the string result for the energy and the eigenvalues 
obtained using the Bethe ansatz 
for the $SO(6)$ chain.
 This also shows that in  this case there is an exact 
 agreement between the energy as 
a function of the conserved quantities as obtained from the solutions 
of our sigma model and the one obtained 
from the Bethe ansatz. This leads us to conjecture that one should be able to 
prove the agreement  in general as was  done in  the $SU(2)$ case in \ci{kmmz}.

As a side  but interesting result,  we also find,  using action--angle variables,  
 an  exact classical relation between 
the energy and the constants of motion of 
 pulsating solutions. 
This simplifies previous treatments,  
putting these  solutions at the same level
 as the rotating ones of \ci{ft2,afrt,art}.


The organization of this paper is as follows.
 In section \ref{SU(3)} we discuss the most general case of
rotating strings in $S^5$ and the gauge fixing  procedure. In the next section 3
 we do the same for the
most general  fast motion on $S^5$
 which includes also pulsating solutions.
 We also explain there
a systematic procedure to compute higher orders in $\tl$ 
 from the string side.
In section \ref{FT} we obtain the same leading order sigma model starting 
from the field theory
side
and
studying the action of the dilatation operator on operators constructed out of scalars, 
i.e. starting  from the $SO(6)$ spin chain  Hamiltonian of  \ci{mz1}.
 We give
examples of the string--spin chain 
 correspondence in the following section 5 
where we  verify that it includes all 
known solutions in which the motion is on the $S^5$. 
Finally, in  section \ref{puls} we give a detailed analysis of the pulsating
solution of \ci{Min2} using action--angle variables.

 In section \ref{conclu} we  make some  concluding remarks,  and
 in  the Appendix we give a  brief introduction
to the subject of canonical perturbation theory to 
make the paper self-contained and as a reference for
the reader.

\section{String theory side: rotating strings ($SU(3)$ sector) \la{SU(3)}} 

%

In preparation for studying the full $SO(6)$ sector, which is done in the next section, 
let us start here by describing the general procedure to derive the  effective action 
for a  rotating string in the limit of large semiclassical rotation parameters. This
connects and generalizes previous partial results of \ci{kru,krt,lopez,ST}.
We shall  isolate and gauge-fix a  ``fast''   collective coordinate 
and get an action for ``slow'' variables 
as an expansion in 
$\tl  = { \l \ov L^2}$, where $L=J$ is the total $S^5$ 
angular momentum.
 This action will thus reproduce 
the expression for the energy  of rotating string solutions 
expanded in powers of $\tl$  \ci{ft2,afrt,art,tse2}. 
In the two-spin 
or ``$SU(2)$''  sector it  will coincide  with the  action found in 
\ci{krt}, while in the most-general pure-rotation 
 three-spin 
or ``$SU(3)$''  sector 
it will  generalize the leading-order action found in \ci{ST,lopez}
(see also \ci{mikk,mikkk}) 
to all orders in $\tl  = { \l \ov L^2}$.
Here we shall explicitly 
consider only the case when string moves on $S^5$
but a generalization to the case when there is also a motion in $AdS_5$ 
is possible too (see \ci{ST} and references there).

 \subsection{Isolating  the ``fast'' angular coordinate $\a$}

In the case of generic rotating strings 
 it is natural 
 to  follow  \ci{ft2,krt,ST} and parametrize the   $S^5$ metric
in terms of 3 complex coordinates $Z_i$ ($i=1,2,3$) 
\bea\la{so}
ds^2= -dt^2 + dZ^*_i  dZ_i \,,  \qquad\qquad Z^*_i Z_i=1\, , 
\eea
where $t$ is the time direction of $AdS_5$.
In terms of 6 real coordinates $X_i$  or standard angles of $S^5$
one has 
$$Z_1  = X_1 + i X_2=\sin \g \ \cos \psi \  e^{i\vp_1 } \ , \ \ \ \  
Z_2= X_3 + i X_4 = \sin \g  \ \sin \psi \ e^{i \vp_2} \ , $$
\be \la{gl}
\ \ \   \  
Z_3= X_5 + i X_6 = \cos \g  \ e^{i \vp_3} \ .  \ee
In this parametrization string states that carry
 three independent (Cartan) 
components $J_i$ of $SO(6)$ angular momentum 
  should be   rotating
in the three orthogonal planes 
 \ci{ft2,tse2}.
To consider the limit of large total spin  $J=J_1 + J_2+ J_3$ we
would like to isolate the corresponding collective coordinate,
i.e. the common phase of $Z_i$.
In the familiar case of  fast motion of the center of mass the role
of $J$ is played by linear momentum or  $p^+$.
Here, however,  $J$ represents
the  sum of an  ``orbital''  {\it and}  ``internal''
angular momenta   and thus
does not correspond  simply to the center of mass motion.
This is thus a
generalization of the limit considered in \ci{bmn}:
we are interested in ``large''
extended string configurations and not in 
nearly point-like strings.

Isolating the common phase  in  the three 
  orthogonal planes by 
introducing the new coordinates $\a$ and   $U_i$ 
\be\la{pot}
Z_i=e^{i\alpha} U_i\,, \ \ \ \ \ \ \ \ 
  \ \ \    U_i^* U_i=1\, ,  
\ee
one finds that the metric \rf{so}  becomes
\ba\la{met}
ds^2= - dt^2 +(d\alpha+C)^2+dU_i^* dU_i
-C^2= - dt^2 +  (D\alpha)^2   + DU_i^* DU_i  \,,
\ea
where
\be\la{mo}
C\equiv -i U_i^*dU_i\,,\ \ \ D\alpha \equiv  d \a + C\,,  
\ \ \  DU_i\equiv dU_i-iCU_i\,, \ \ \ DU_i^*\equiv dU_i^* + i C U^*_i. 
\ee
Here $U_i$  belongs to $CP^2$: 
the metric is invariant under a simultaneous shift 
of  $\a$ and a rotation of $U_i$. In general, 
this parametrization corresponds to a Hopf $U(1)$ fibration of $S^{2n+1}$ 
over  $CP^{n}$:  
$ds^2 =  DU_i^* DU_i$ is the Fubini-Study metric 
and $K=\ha dC$ is the covariantly constant  K\"ahler form on 
$CP^n$.
In the two-spin or $SU(2)$ sector ($Z_3=0$) 
 where the motion is within
$S^3$ of $ S^5 $  the two  coordinates $U_r$  of $CP^1$ can be replaced 
by a unit 3-vector \ci{krt}
\be \la{nen}
n_i \equiv  U^\dagger \sigma_i U \ , \ \ \ \ \ \ \ \
U=(U_1,U_2) \ , \ \ \ \  \ \ 
D U^*_r D U_r= \fo   dn_i dn_i \ ,  
\ee
and $C$ has a non-local WZ-type representation
$
 C = -\ha  \int^1_0  d \xi\  \epsilon_{ijk} n_i  \del_\xi
  n_j  d n_k.$

The general form of the string  action in $R_t \times S^5$ 
(the string is positioned at the center of $AdS_5$ with 
$t$ being the $AdS_5$ time) 
with the metric \rf{met} 
is then (we use world-sheet  signature $(-+$))
\be \la{ste}
I= { \sql} \int d\tau \int^{2\pi}_0  {d \s\ov 2\pi}  \ \LL \ , \ee
\be \la{st}
\LL =  - \ha \sqrt{ - g} g^{pq}
\big( - \d_p t \d_q t + D_p \a  D_q \a  +  D_p U_i^* D_q U_i
\big)  \ . \ee
The crucial point is that 
 one  should view  $t$ and $\a$ as ``longitudinal'' coordinates
that reflect
the redundancy of the reparametrization-invariant
string description: they are not ``seen''
on the gauge theory side, and should be gauged away (or
eliminated using the constraints).
At the same time,  the $CP^2$ vector  $U_i$ 
describing string profile 
should be interpreted as
 a ``transverse'' or physical
 coordinate which should thus  have a counterpart
 on the  spin chain side, with an  obvious candidate
 being  a vector parametrizing the coherent state \ci{kru,krt,ST}.
 The conserved  charges corresponding to translations
 in time and   $\a$ are 
 \be\la{chal} E= \sql \E \ , \ \  \ \ \ \ \ \
\E= - \inti \sqrt{ - g} g^{0p}\del_p t \ , \ \ \
\ee
\be  \la{kop}
J= \sql \J \ , \ \  \ \ \ \ \ \ \J=\inti\  P_\a \ , \ \ \ \ \   \ \   \ P_\a = 
 -   \sqrt{ - g} g^{0p} D_p \a  \ ,  \ee
 where 
 the effective coupling constant $\bl$  is
directly related to the (rescaled) 
charge  $\J$ in \rf{kop}
\be\la{lama}
\bl \equiv {\l \ov J^2} = { 1 \ov \J^2} \ ,  \ \ \ \ {\rm i.e.} 
\ \ \ \ \ 
\J = { 1 \ov \sqrt{\bl} } \ .  \ee
The ``fast motion'' expansion in powers of $1\ov \J^2$  which we 
will be interested in is, thus, the 
 same as the  expansion in powers of $\bl\to 0$.

 What remains is to do the following  three steps:
 (i) proper gauge fixing; (ii) expansion of the action at large $\J$, 
 suppressing time derivatives of $U_i$ in favor of 
 spatial derivatives;
 (iii) field redefinitions to eliminate from the expanded action 
terms of higher order than first in time derivatives. 

 To leading order in $\tl$ one may simply use the standard 
 conformal gauge (as was done in the  $SU(2)$
 sector in \ci{kru,krt} and in the  $SU(3)$ sector in \ci{ST,lopez}).
 As was 
 pointed out  in \ci{krt}, to get the full action to all orders in $\tl$ 
 one  should use a  special ``adapted''  gauge.
 Here we shall make this first gauge fixing step particularly transparent 
 by  explaining that 
 the gauge fixing procedure used in \ci{krt}
 amounts simply to the standard static gauge for  the 
 coordinate which is 
  2-d dual (or ``T-dual'') to  $\a$. 
 
Having in mind comparison
with the spin chain side  it is natural to request that
translations in time in the target space and on the
world sheet should 
be related. Also, we should ensure that  the angular momentum $\J$
 is
homogeneously distributed along the string  so that 
its density $P_\a$  in \rf{kop}, i.e. the  momentum conjugate to $\a$,
 is constant.
 Therefore, one should   fix the following gauge \ci{krt}
\be \la{gaug}
 \ \ t= \tau \ , \ \ \ \ \ \ \ \  \ \ \ \ 
 \ \  P_\a = \J =\const \ . \ee
As was shown in  \ci{krt}, 
starting with the phase-space form of the string action and 
imposing   this ``non-conformal'' 
 gauge  
one finds the following effective 
Lagrangian for $U_i$ 
\be \la{klpo}
\LL  = \J C_0    -  \sqrt{ - \det\  \tid h_{pq} } \ ,\ee
\be  \la{kla}
\tid h_{pq} = \tid \eta_{pq} +  D_{(p} U^*_i D_{q)}  U_i \ , \ \ \ \ \ \ \ 
\tid \eta_{ab} \equiv  {\rm diag} ( -1, \J^2) \ , \ \ \ \ \ \ 
C_p= - i U^*_i \del_p U_i \ .   \ee
In the $SU(2)$ case \ci{krt} this gives 
an  equivalent action for 
 $n_i$ \rf{nen} with 
$\tid h_{pq} = \tid \eta_{pq} + \fo \del_p n_i \del_q n_i $.
As was noted in \ci{krt}, 
apart  from  the WZ-type  term $C_0$,
 the Lagrangian 
\rf{klpo}   looks like a 
Nambu Lagrangian in a static gauge, 
 suggesting that there may be
 a more direct way of deriving  it. 
 This is indeed the case as we shall  explain below.

 \subsection{2-d duality transformation $\a \to \tilde \a$ 
and ``static'' gauge fixing}

Let us make few   remarks on 
 interpretation of string action in connection with spin chain 
on the SYM side 
 and for simplicity   consider the $SU(2)$ case. 
In the  semiclassical coherent state description 
of the  spin chain \ci{kru}, one has  a circular  direction 
along the  chain  at each point of which one has a 
classical spin vector  $\vec n$ belonging to a  2-sphere. 
In other words, if we 
combine  operator ordering direction under the trace 
 with an ``internal''  direction
we get  $S^1 \times S^2$  type of geometry.
A  similar geometry indeed 
  emerges on the  string sigma model side -- we have 
$S^3$ fibered  by $S^2$ with base   $S^1$. 
However, in \rf{met} the  $S^1$  direction is that of  the angle $\a$.
Fast rotation corresponds to time-dependent $\a \sim t + \ldots$, so 
it is not quite  appropriate 
 to call $\a$  a  ``longitudinal'' coordinate  since after 
we have  chosen  $t =\tau$ gauge we  have 
already  fixed  a time-like coordinate. 
A natural  idea is that the true longitudinal coordinate 
should be ``T-dual''   counterpart $\ta$ 
of $\a$, i.e. one should apply 2-d duality to 
the scalar field $\a$ in \rf{st}.\foot{We are grateful
to   N. Nekrasov for emphasizing
to us  the potential importance   of applying  T-duality in $\a$.}

There is, however, an important subtlety. 
The standard discussions of  T-duality are usually done
in conformal  gauge, 
but if   we would fix the  conformal gauge and then also  $t=\tau$ 
we would  no longer have a  freedom to  fix $\ta \sim \s$. 
Actually, $\ta \sim \s$ will not, in general, be a solution of the
equations for $\ta$ in the conformal gauge.
The correct procedure is not to impose the  conformal gauge;
 we should first apply  the 2-d duality, then go from the Polyakov 
to
the  Nambu form of the action by
 solving for the  2-d metric $g_{pq}$, 
and  finally fix the static gauge 
 $t \sim \tau,\  \ta \sim \s$. 
Remarkably, this turns out to be {\it equivalent}
 to the gauge fixing procedure used in \ci{krt}, leading  directly to  
\rf{klpo}.  This  explains that the 
 non-diagonal gauge  used in \ci{krt}   is nothing but 
  the standard {\it static} gauge in the Nambu action for 
 the {\it dual} coordinate 
$\ta$.
 Not imposing the 
conformal gauge allows one to have solutions 
consistent  with the above static gauge choice.


Let us first  note that the equation for $\a$ following
 from the action \rf{st}, i.e.
 $\del_p (  \sqrt{ - g} g^{pq} D_q \a)$ =0,    
 can be  solved by setting 
\be \la{hohu}
\sqrt{ - g} g^{pq} D_q \a = -   \ep^{pq} \del_q \ta \ , 
\ee 
where   $\ta$ should then satisfy 
\be   \la{alp}
\del_p (  \sqrt{ - g} g^{pq} \del_q \ta)
+ \ep^{pq} \del_p C_q =0 \ , \ \ \ \ {\rm i.e.} \ \ \ 
\del_p (  \sqrt{ - g} g^{pq} \del_q \ta) 
 =i  \ep^{pq}   \del_p U^*_i \del_q U_i \ .   
\ee
Comparing \rf{hohu} to \rf{kop}  we observe that 
\be 
\J =    \inti \ \del_1 \ta  \ ,\ \  \ \ \ {\rm i.e.} \  \ \ \ \ \ 
\ta (\tau, 2\pi) = \ta (\tau,0) + 2 \pi  \J \ ,  \ee  
which is satisfied, in particular, if one  
fixes the gauge  by setting  
\be  \la{sto}  \ta  = \J \s   \ . \ee 
The limit of small $\bl$ or large $\J$ (cf. \rf{lama}) 
is then  the limit of large winding number of the dual coordinate $\ta$.
\foot{A comment on the quantization condition. In the standard case 
of circular coordinate with radius  $a$  the dual coordinate 
has radius $\tid a = {\a' \ov a^2}$. In the present case 
$\a' = { 1 \ov \sql}$ and $a=1$ so  the period of 
$\ta$ should be  ${ 2 \pi \ov \sql}$. This implies 
that $J = { \J \ov \sql}$ should  indeed be an integer winding number.}

Let us now  apply the  2-d duality systematically  at the level 
of  the string action 
\rf{st}.  Replacing $D_p \a$ by $A_p + C_p$  where $A_p$ is an auxiliary
2-d vector field, adding  the  ``Lagrange multiplier'' term 
$\ep^{pq} A_p  \del_q\ta $, and then integrating out $A_p$   we end up
with the 2-d dual counterpart of \rf{st}
\be \la{stt}
\LL =  - \ha \sqrt{ - g} g^{pq}
\big( - \d_p t \d_q t + \del_p \ta  \del_q \ta  + D_p U^*_i D_q U_i
\big)  +   \ep^{pq} C_p \del_q \ta\   ,  \ee
where $C_p$ was given in \rf{kla}. 
Thus the ``T-dual''  background  has no off-diagonal
 metric component 
but has a  non-trivial NS-NS 2-form  coupling 
in the $(\ta,U_i)$  sector. 
Eliminating the  2-d metric $g^{pq}$ we then get the Nambu form of the 
2-d dual counterpart of the action
 \rf{ste},\rf{st}   
\be \la{sett}
\LL  =     \ep^{pq} C_p \del_q \ta\  - \sqrt{  h } \   ,  \ee
\be \la{ggm} 
h =|\det \  h_{pq}|\ , \ \ \ \ \ \ 
h_{pq} = 
- \d_p t \d_q t + \del_p \ta  \del_q \ta 
 + D_{(p}  U^*_i D_{q)} U_i\ . 
\ee
If we now   fix the  static gauge 
\be \la{stop}
t = \tau \ , \ \ \ \ \ \ \ \ \  \ta= \J \s  \ , \ee 
we finish with the action equivalent to 
\rf{klpo},\rf{kla} 
\be 
\la{eaa}
I =  J \int dt \inti \ \LL \ , \ \ \ \ \ \ \ \ \ \ \ \ \ 
\LL  =  C_0  -  \sqrt{  \tid h  }  \    ,  \ee
\be \la{ity}
\tid h = (1+  \tl  |D_1 U_i|^2)( 1 -  |D_0 U_i|^2  )
+  \fo \tl   (D_0 U^*_i D_1 U_i + c.c.)   ^2 \ . 
\ee
We thus uncover  the origin of the string-theory  
counterpart   of the WZ term $C_0$  in the spin-chain
coherent state  effective 
 action:
it comes from the 2-d  NS-NS WZ term upon
the  static gauge fixing in the T-dual  $\ta$ action.

 \subsection{Eliminating time derivatives }

The remaining steps are as  in \ci{kru,krt}.
We assume that higher powers of  time derivatives are suppressed.
To  define  a consistent ${1\ov \J^2}=\bl $ expansion,
we may then  redefine the  time  coordinate so that the leading order 
approximation does not involve $\bl$: 
\be\la{resc}
\tau \to \J^2 \tau = { \bl^{-1}} \tau\ , \ \ \ \ {\rm i.e.}\ \ \
\ t \to   \bl^{-1} t \ , \ \  \ \ \ \ \
\del_0 \to\  \tl\  \del_0 \ , \ee
thus getting the string action  \rf{eaa} with 
\be \la{bbb}
\LL=  C_0 -  \tl^{-1}  
\sqrt{ (1+  \tl  |D_1 U_i|^2)( 1 - \tl^2 |D_0 U_i|^2  )
+  \fo \tl^3   (D_0 U^*_i D_1 U_i +c.c.)^2} \ .  
\ee 
Expanding in powers of $\tl$ 
 (and  omitting the  constant term)
this gives\foot{In the $SU(2)$ case one finds  \ci{krt}
$\LL= C_0 - { 1 \ov 8  } (\d_1 n_i)^2  +  {\tl  \ov 8 }
\big[(\del_0
n_i)^2   +  { 1 \ov 16}
(\del_1 n_i)^4  \big] $ $ 
- \  {\tl^2 \ov   32} \big[
  (\del_0 n_i \del_1 n_i)^2
-{1 \ov 2}   (\del_0 n_i)^2  (\del_1 n_k)^2
+ { 1 \ov 32 } (\del_1 n_i \del_1 n_i )^3  \big]
 + O(\tl^3)$.} 
\be \la{accv} \LL=\LL_1 + \tl \LL_2 +  O(\tl^2) \ , \ \ \ \ \ \ \ \ \ \ \ 
\LL_1 = - i U^*_i \del_0 U_i  - \ha |D_1 U_i|^2 
\ , \ee \be \la{subb}
\LL_2=  \ha  |D_0 U_i|^2  + {1 \ov 8} |D_1 U_i|^4     \ .  \ee 
Finally, we should 
eliminate higher than first 
 time derivatives in $L$ by field redefinitions 
 order by order in $\tl={1\ov \J^2}$. This was explained in detail 
in the $SU(2)$ case in \ci{krt}.
For example,  this amounts to eliminating $\del_0 U_i$ from $\LL_2$ 
using leading-order  (Landau-Lifshitz  type) 
equation for $U_i$ following from $\LL_1$.
Note that the  leading  term $\LL_1$ in the action \rf{accv}
 is  the same as found by choosing the conformal gauge and eliminating
 $\a$ from the action using the constraints \ci{ST}. 
Also,  in the next section  we 
will again obtain  an   equivalent  result by a 
different but related method which relies 
on the use of canonical transformations.


To summarize, the  main lesson of the above derivation 
 is that 
the dual angular  coordinate $\ta$ is just a replacement for the momentum
$P_\a $ 
of the ``fast'' coordinate $\a$. 
Applying T-duality $\a \to \ta$ 
 allows one to fix the static gauge $\ta \sim \s$ 
 which thus identifies the spin chain direction with $\ta$ direction.
Applied to $SU(3)$ sector the above procedure  determines the 
subleading terms in the effective action.
$\LL_2$ in \rf{subb}  may, in principle, 
be compared with the 2-loop effective action on the spin chain side
generalizing the comparison in the $SU(2)$ case  in \ci{krt}.
Indeed, the  corresponding  2-loop dilatation operator 
(for the $SU(3|2)$ superspin chain containing bosonic $SU(3)$ sector)
was found in \ci{beit}, and, 
as was recently pointed out in \ci{Min3}, 
the mixing of bosonic operators from the $SU(3)$ sector 
with the fermionic operators at  the  
 two (and higher) loop order is effectively suppressed  
in the long spin chain (large $L=J$) limit.

\section{String theory side: general fast motion  
  ($SO(6)$ sector) \la{SO(6)}} 

 Now we can proceed to analyze the $SO(6)$ sector.
 On the string side we have a string that moves almost at the
speed of light\footnote{The corresponding 
 $\tl \to 0$ limit  can also be seen as
 a small tension limit \ci{mateos}.} 
along the $S^5$. It is then natural  to start by 
isolating a coordinate that describes the fast motion in order to use 
the approximation that the velocities of all {\it other}
 coordinates 
are small. In previous work
\ci{kru,krt,lopez,ST}\footnote{See also \cite{mikk} for a
 different approach.} 
 this was achieved  by means of an appropriate change of coordinates.
However, this can  be done only 
if we already know the particular string configuration 
 we are aiming to describe.
Here, instead,  we would like  to  isolate  a  fast coordinate 
 independently of the type of solution, i.e. 
 in  a universal way that will apply to  
rotating \ci{ft2,ft4,bfst}, pulsating \ci{Min1,Min2} and 
other similar solutions (see \eg\ \ci{dev,Dim}) that 
describe fast moving strings on  $S^5$.
    
 The common feature to all of them is that each piece of the string is
 moving along a maximum circle 
almost at the speed of light.   Since each point of the string moves,
 in general, along  a 
different massless  geodesic in $R_t \times S^5$, to define 
 the fast coordinate we need to know the 
position {\it and} velocity of each point of the string.
 Therefore,
we can isolate the fast coordinate only 
if we use the {\it phase space} description.

 What we find is that the circle along which each piece of string moves,
 slowly changes in time
due to interactions with its neighbors along the string (represented
 by the terms containing sigma derivatives). 
This is in exact agreement with the picture on the spin chain side. In an appropriate gauge, 
each piece of string moving along
a circle carries one unit of R-charge and corresponds to a half-BPS operator on the spin chain
carrying the same charge. On the spin chain, the operator
 we have at each site (given, for example, 
 by the mean value of the spin in the $SU(2)$ case) 
also changes in time precisely as a result of the 
interaction with its neighbors.
 Notice that here we are mapping a time dependent classical string into a 
time dependent coherent  
state on the spin chain side and not 
into an energy eigenstate.
Energy eigenstates of the spin chain or SYM theory  should correspond to single-string 
eigenstates in the bulk, while classical string solutions should be 
represented by coherent states.

In the rest of this section we do the following steps.
 First, we isolate the fast coordinate which we 
shall call $\alpha$ as in section 2. 
Then we obtain the leading order terms in the action 
by considering the limit in which $\alpha$ changes much faster than all 
 other 
coordinates. One problem is that the Lagrangian contains terms which oscillate in $\alpha$ (proportional 
to $e^{\pm2i\alpha}$).
 Since they average to zero, at leading order they can be discarded. However, when we go to higher
orders, they give a contribution. The simplest way to treat them is to  eliminate them order by order 
using coordinate transformations. In this way we end up with a Lagrangian that, to the order considered, 
is independent of the fast coordinate, i.e., like \rf{st}, has 
 an isometry  $\alpha\rightarrow \alpha + \mbox{constant}$. 
We can then perform a T-duality $\alpha\rightarrow\tilde{\alpha}$ and fix a static 
gauge  as in \rf{sto},\rf{stop}, i.e. 
$\tilde{\alpha}= \bL \sigma$, 
 where $\bL={L \ov \sql}  $ is essentially the momentum conjugate to $\alpha$.
 We end up with the  standard Nambu action with an extra term due to the presence 
of a $B_{mn}$--field. This  
Wess-Zumino type term  ensures the correct phase space structure of the action. 
After that the large energy  or large $L$ expansion 
is simply a Taylor expansion of the square root in the Nambu action.
 
 \subsection{Isolating the ``fast'' angular coordinate $\a$}

Consider the Lagrangian for the string on $R\times S^5$  
written in terms of $AdS_5$  time $t$ and  6 real  coordinates  $X_m$
\be \la{lap}
\LL = - \ha [- (\del_p t)^2 + (\del_p X_m)^2  + \Lambda ( X_m X_m -1)]  \  , 
\ \ \ \ \ \ \ \  m=1,\ldots,6 \ , \ \ p=0,1\ .  \ee
  The  corresponding conformal
 gauge constraints  are (we  choose $t=\k \tau$ as an additional gauge fixing
condition)
 \be \la{cob}
(\del_0 X_m)^2 + (\del_1 X_m)^2 = \k^2 \ , \ \ \ \ \ \ \ 
\del_0 X_m \del_1 X_m = 0 \ . \ee 
In the simplest case  when $X_m$ does not depend  on $\s$, i.e. the string
is point-like,  we get the geodesic equations  
$ \del^2_0 X_m  + \Lambda   X_m =0$, $\Lambda= (\del_0 X_m)^2= \k^2$.
These  are  solved by 
\be \la{geo}
X_m(\tau)  = a_m \cos \alpha  + b_m \sin \alpha \ , \ \ \ \ \  \ \alpha = \k \tau \ , \ \ \ \ 
a^2_m=1 \ , \ \ b_m^2=1 \ , \ \ \ a_m b_m =0  ,  \ee
or,  equivalently,  by 
\be \la{fas}
X_m =  \frac{1}{\sqrt{2}} \left(e^{i\alpha} V_m + e^{-i\alpha} V^*_m \right) \ , \ \ \ \ \ \ \ \ \ \ \ \ \ 
\ V_m = { a_m - i b_m  \ov \sqrt 2} \ , \ee 
where 
\be \la {kopl}
V_m V^*_m = 1 \ , \ \ \  \ \ \ V_m V_m  =0 \ ,
 \ \ \ \ \ \ \  V^*_m V^*_m=0 \ . \ee
The constant 6-vectors $a_m$ and $b_m$ parametrize the space of massless 
geodesics in $R \times S^5$, \ie\ the space of maximum circles in $S^5$. 
Equivalently, they
parameterize the Grassmanian $G_{2,6}$ --  the space of planes in $R^6$ that go through the 
origin.\footnote{In general, the Grassmanian $G_{k,n}$ is the space of all $k$ dimensional 
hyperplanes 
of the vector space $R^{n}$.}
Thus, as was  noted   in \ci{mikk}, 
the coset  $G_{2,6}=SO(6) /(SO(4)\times SO(2))= SU(4)/S(U(2) \times U(2)) $ 
is the moduli space of geodesics in  $S^5$. 
On  the SYM side the same space   appeared 
as a natural coherent state  space  for $SO(6)$ spin chain in  \ci{ST}.
 Let us mention also 
that $V_m$ describing the Grassmanian 
  may be interpreted as  coordinates of $CP^5$ subject
 to an additional condition $V^2_m=0$, i.e. 
they define an 8-dimensional subspace  of $CP^5=SU(6)/U(5)$.

 If  we now consider an extended relativistic 
closed string, the trajectory 
of each piece of the string can be described 
as a circular trajectory but with the parameters $a_m$ and $b_m$ (or $V_m$) 
which determine its orientation, 
slowly changing in time $t=\k \tau$ 
 and  in $\sigma$ (the coordinate along the string). Also, 
 the speed 
along the circle is time dependent in this case. 
A familiar analogy  is an  orbital motion of a planet in a solar system.
 It moves along  an ellipse around the Sun but the parameters of the ellipse slowly 
change in time due to perturbations by other planets, moons, etc.
 In both cases, the fast coordinate is an angle in the plane of motion 
determined by the position 
and the velocity. 

 To obtain the leading order result for an  effective  Lagrangian 
of ``slow'' coordinates 
 it is sufficient 
 to work in the  conformal gauge and use again  the additional 
gauge condition  $t=\kappa\tau$.
Then   the phase space Lagrangian is simply
\beq \la{lak} 
\LL = -\half\kappa^2 + P_m \dot{X}_m - \half P_m P_m - \half X'_mX'_m - \half \Lambda (X_mX_m-1)\ . 
\eeq
Here and below we define the momenta 
($P_m$, etc.) as derivatives of the Lagrangian, i.e.
do not include the tension $\sql$ factor in the action \rf{ste} 
in their definition. 
To isolate the coordinate $\alpha$ we introduce its conjugate momentum 
$P_{\alpha}$ and do the following 
coordinate transformation
\beqa
 X_m &=& \cos\alpha\ a_m + \sin\alpha\ b_m \ ,  \non\\
 P_m &=& P_{\alpha} (-\sin\alpha\ a_m  +  \cos\alpha\ b_m) \ , 
\label{pht}
\eeqa
where, as in \rf{geo},
 $a_m^2=1$, $b_m^2=1$, $a_mb_m=0$.
 This can be  conveniently written as\footnote{For constant $\a$, this is actually quite analogous 
to the usual transformation 
$(q,p)\rightarrow(a,a^*)$ to ``coherent coordinates''   for a harmonic oscillator.}
\beqa 
X_m &=& \frac{1}{\sqrt{2}} \left(e^{i\alpha}V_m+e^{-i\alpha}V^*_m  \right)\ , \non \\
P_m &=& \frac{1}{\sqrt{2}}i P_{\alpha}\left(e^{i\alpha}V_m-e^{-i\alpha}V^*_m  \right)\ \la{xp} , 
\eeqa
where as in \rf{kopl} 
\beq
V_m = \frac{a_m-ib_m}{\sqrt{2}}\ , \ \ \ \ \ \ \  V_m V^*_m =1 \ , \ \ \ \ \ \ V^2_m =0 \ . 
\eeq
The constraints on $V_m$  ensure that $X^2_m =1$ and 
$P_m X_m =0$ (which is a consequence  of $X^2=1$ and $P_m= \dot{X}_m$ in the conformal gauge). 
There is a redundant degree of freedom which is obvious from  the gauge invariance
($\beta=\beta(\sigma,\tau)$ is an arbitrary function):
\beq\la{gaugeinv}
 \alpha \rightarrow \alpha-\beta\ , \ \ \ \ \ \ 
\ V_m\rightarrow e^{i\beta}\,V_m\ ,\ \ \ \ \ V^*_m\rightarrow e^{-i\beta}\, V^*_m\ .
\eeq
 We could fix this gauge invariance by 
choosing \eg\ $a_6=0$ or $V_6 + V^*_6=0$, 
but it is more convenient not to do this 
 to preserve the $SO(6)$ rotational symmetry. 
It is then natural to 
introduce the covariant derivatives defined by 
\beq \la{covdev}
D_p \alpha = \partial_p\,\alpha + C_p\ ,\ \ \ \ D_p V_m = \partial_p V_m - iC_p V_m\ ,\ \ 
\ \ \ D_p V^*_m  = \partial_p V^*_m + i C_p V^*_m \ , 
\eeq
where as in \rf{mo}  the $U(1)$ 
gauge field $C_p$ is not an independent variable but is given by 
\beq
C_p = -i V^*_m \partial_p V_m\ . 
\eeq 
Now it is easy to derive the following useful 
identities\foot{
We shall often omit index $m$ on $V_m$ in quadratic relations 
 assuming summation over $m$.}
\beqa
&&P_m \partial_p X_m   = P_\alpha D_p\alpha\ ,\ \ \ \ \ \ \ \ \ \ \ \ \ 
P^2_m  = P_{\alpha}^2 \ , \non \\
&&X'^2_m  = (D_1\alpha)^2 + |D_1V|^2
 + \half\left[e^{2i\alpha} (D_1V)^2 + e^{-2i\alpha}(D_1 V^*)^2\right]\ , 
\eeqa
which allow us to rewrite the Lagrangian \rf{lak} as 
\beqa
\LL &=& -\half \kappa^2 +  P_\alpha D_0\alpha -\half  P_{\alpha}^2 
-\half \left\{(D_1\alpha)^2 + |D_1V|^2 
 + \half\left[e^{2i\alpha} (D_1V)^2 + e^{-2i\alpha}(D_1V^*)^2\right]\right\} \nonumber\\
&& - \mu_1 (|V|^2 -1) -\half \left(\mu_2 V^2+c.c.\right)  \ , \la{bv}
\eeqa
where $\mu_1$ is a real and $\mu_2$ is a complex Lagrange multiplier field. 
The conformal gauge  constraints are then 
\beqa
&&P_mP_m+X'_mX'_m =  P_{\alpha}^2 
  + (D_1\alpha)^2 + |D_1V|^2 + \half\left[e^{2i\alpha}
 (D_1V)^2 + e^{-2i\alpha}(D_1V^*)^2\right]=\kappa^2\ , \nonumber\\
&& P_m X'_m = P_\alpha D_1\alpha = 0 \ .\la{cvb}
\eeqa
Notice that we could use that $(D_1V)^2 = (\partial_1V)^2$
 (since $V^2=0$ and $V\partial_1V =0$) 
to simplify these expressions 
but we prefer to keep  the $U(1)$ gauge invariance \rf{gaugeinv}  manifest. 

 Another point we note for later reference is that the second 
constraint in (\ref{cvb}) 
implies $D_1\alpha=\partial_\sigma \alpha + C_1 =0$.
Since the string is closed, the coordinates in \rf{xp}  must be periodic in
$\s$. We may assume then that $\alpha (\tau, \s+ 2 \pi)
= \alpha (\tau, \s)$ (an a priori possible ``winding''  part can be absorbed into 
$V_m$, cf. \rf{gaugeinv}). Then  
 we obtain the following constraint 
\beq 
0 = \int_0^{2\pi} \frac{d\sigma}{2\pi} \partial_\sigma \alpha = - \int_0^{2\pi}  \frac{d\sigma}{2\pi} C_1 = 
i  \int_0^{2\pi} \frac{d\sigma}{2\pi} V^*_m\partial_\sigma V_m \ . 
\la{P=0}
\eeq
 On the field theory  side,  the fact that the string is closed is
 related to the fact that the dual 
operators are  given by the single  $SU(N)$  trace
 which is invariant under 
cyclic permutations of local fields under the trace.
In the  spin chain description
such operators are represented by states  of a closed chain which are 
 invariant under cyclic permutations.
This invariance gives rise to  a condition equivalent to (\ref{P=0})
(see section 4).

\subsection{Leading order in large energy expansion}
  
We can take into account that $\alpha$ is a fast variable by
 setting 
$$  \alpha = \kappa \tau + \bar{\alpha} \ , \ \ \ \ \ \ \ \ \ \ 
P_{\alpha}=\kappa +p_{\alpha}
\ , $$
where $\bar{\alpha}$ and $p_\a$ are the new variables, 
and then taking  the limit
$\kappa\rightarrow\infty$ while 
keeping $\kappa \dot{V}$, $\kappa\dot{\bar \alpha}$ and $\kappa p_{\alpha}$ 
fixed.\footnote{See \ci{mik,mikk} for related ideas. 
This limit is also reminiscent of the so called wrapped or
non-relativistic limit \ci{wrapped}.}
The only difference with the  previous 
rotating string 
cases is that here there are terms in \rf{bv},\rf{cvb}  proportional 
to $e^{\pm i\kappa\tau}$. It is
clear that for large $\kappa$ these terms
 can be ignored since they average to zero.\footnote{An equivalent averaging  
was also considered in \cite{mikk}.} 
Then,  the conformal gauge constraints \rf{cvb} 
 determine $\bar \alpha'=\a'\ (\equiv\partial_1 \a)$ and $p_\alpha$ as
 (to leading order in $1\ov \kappa$): 
\beq\la{huo} 
D_1 \alpha=0\ \ \Rightarrow\ \  \alpha' = -C_1= i V^* \del_1 V  \ , \ \ \ \ \ \ \ \ \ 
\kappa p_\alpha = -\half D_1V D_1V^* \ . 
\eeq
Using that  $D_1\alpha=0$ in the Lagrangian \rf{bv}  and taking the same 
$\k \to \infty$ limit we find 
\beq
\LL = \kappa\dot{\bar{\alpha}} - i  \kappa V^* \dot{V}  - \half D_1VD_1V^*  
  -\mu_1 (|V|^2-1) - \half (\mu_2  V^2   + c.c.) \ .
\label{hyy}
\eeq
The first term here is a total derivative and may be omitted, so that 
the corresponding action  takes the form similar to \rf{eaa},\rf{accv}
\be \la{vbvb}
I =  L  \int dt \inti \ \LL_1  \ , \ \ \ \ \ \ \ 
L= \sql \bL \ , \ \ \ \ \bL \equiv {1 \ov {\sqrt{\tilde \l}}} 
 \simeq \k \simeq  P_\a \ , \ee
\be\la{ji}  \LL_1 = - i  V^* \dot{V}  - \half |D_1 V|^2  \ , \ee 
where in \rf{ji} we  rescaled the time coordinate as in \rf{resc} (and did 
not explicitly include the Lagrange multiplier terms).

We thus find the ``Landau-Lifshitz''  version of the 
 $G_{2,6}$  sigma model, i.e. 
 with the first time derivative WZ-type  term instead of the usual 
 quadratic time-derivative term.\foot{Note that this  model is  integrable
 since it is obtained as a limit of
an integrable $SO(6)/SO(5)$ sigma model (in general, 
Grassmanian  cosets are classically  integrable \ci{ei}).}
This  Lagrangian   \rf{hyy} is the same 
as found in  \ci{mikk}  through a different
procedure.\foot{One  difference compared to \ci{mikk}
is that we write $\LL$ in $U(1)$  gauge invariant form. 
The expression in \ci{mikk} should be considered as a 
gauge fixed version  of \rf{hyy}.}

A Lagrangian  equivalent to \rf{hyy} was found also 
on the spin chain side 
in \ci{ST} (see also the next section for the  derivation). 
As was shown in \ci{ST}, 
the corresponding Hamiltonian 
 represents a  low-energy (continuum) limit of the 
expectation value of the $SO(6)$ spin chain Hamiltonian of  \ci{mz1} 
in the natural $SO(6)$ coherent state $\ket{\V}$ obtained by applying 
the $SO(6)$  transformation from $G_{2,6}=SO(6)/[SO(4)\times SO(2)]$ 
to the  BPS ground  state  $\ket{0}=(0,0,0,0,1,i)$
at each site of the spin chain. 
This state may be parametrized by a real $6\times 6$ antisymmetric 
 matrix m$_{mn} 
= \bra{\V} M_{mn} \ket{\V}$, where $M_{mn}$ is a hermitian 
 $SO(6)$ generator, 
satisfying the constraints \ci{ST} m$^3=$m,\ \  tr(m$^2)=2$. 
 We can see the relation to the present discussion 
 if we solve these constraints 
by introducing a complex 6-vector $\V_m$ subject to $|\V|^2=1,
 \ \V^2=0$, with 
m$_{mn}=i( \V_m \V^*_n - \V_n \V^*_m)$. 
Going back to the string theory side, 
let us note  that the $SO(6)$ generators corresponding to the Lagrangian 
 \rf{lap} 
  written in terms of 
$X_m$  become, after  using \rf{xp},
\be
M_{mn} = \sql \inti ( X_m P_n - X_n P_m)
\simeq
i  L \inti ( V_n V^*_m - V_m V^*_n) + \ldots \ ,  
  \la{MV} \ee
where we used that $L \simeq  \sql  P_{\alpha}$.
This is the  same result as found 
for the coherent state expectation value of 
the rotation generator  on the spin chain side \ci{ST}, 
implying the identification between
the corresponding complex vector variables $\V_m$ and $V_m$. 

The equivalence between the string and spin chain effective actions 
explains and generalizes the previous results 
about the agreement between the 
leading-order terms in the  energies of  pulsating/rotating strings 
and dimensions of the corresponding gauge theory operators
 which were found  using 
the  Bethe ansatz  approach \ci{Min1,bmsz,Min2,Min3}. 
In addition to its universality, a bonus of the coherent state approach 
\ci{kru}, on which we shall elaborate below, 
is that it explicitly  relates the form of  the string solution 
to the structure  of the  corresponding 
coherent 
 operator on the gauge  theory side. 
We shall  further clarify and illustrate this 
point in  sections 4 and  5  by considering particular  examples
of pulsating solutions. 
But first 
let us generalize the above discussion  and 
 show how we can  obtain, in a systematic fashion, 
higher order corrections to the action \rf{vbvb}.

\subsection{Elimination of the fast variable and higher order terms}

If we go to higher orders, working in 
the conformal gauge introduces a problem pointed out in \ci{krt}: in the present case 
it is related to  the fact that $P_{\alpha}$
 is  identified (to leading order) with $\kappa={E\ov \sql}$. 
The expansion of the
energy will therefore contain powers of the energy itself.
 We would prefer to associate 
$P_{\alpha}$ to a conserved momentum (like $J$ in the rotating case or $L$ 
in the pulsating one). 
While in the conformal gauge this is the case at   the leading
order this is not so  at the higher orders of large energy  expansion. 
Furthermore,  we want such conserved momentum to be uniformly distributed 
along the string (i.e. do not depend on $\s$)  so that we can establish  a simple 
correspondence with the spin chain side.  

The solution to this problem is the same as in \ci{krt}. 
 Instead of fixing  the conformal gauge as we did  in the previous subsection,
we can use   the phase space description 
without fixing the world-sheet metric $g_{pq}$. Then, fixing the partial gauge  
$t=\tau$ (here we prefer  not to  explicitly  include the factor of $\k$), we 
 get the following phase space Lagrangian (cf. \rf{lak}):
\beq
\LL = -P_t + P_m\dot{X}_m - \frac{1}{2\sqrt{-g}g^{00}} \left(-P_t^2 + P_mP_m + X'_mX'_m\right)
 -\frac{g^{01}}{g^{00}} P_m X'_m -  \half \Lambda \left( X_mX_m-1\right) \ .
\eeq
 Doing the same transformation as in (\ref{xp}) we
obtain the Lagrangian in terms of  the new variables (cf. \rf{bv}) 
\beqa
\LL &=& -P_t + P_\alpha D_0\alpha - \frac{1}{2\sqrt{-g}g^{00}} \bigg(-P_t^2 
   +P_{\alpha}^2 + (D_1\alpha)^2 + |D_1V|^2  \nonumber  \\
  && + \ \half\left[e^{2i\alpha} (D_1V)^2 + e^{-2i\alpha}(D_1V^*)^2\right]
\bigg)
 -\frac{g^{01}}{g^{00}} P_\alpha\, D_1\alpha \nonumber  \\
 && -\  \mu_1 (VV^* -1) -\half \left(\mu_2 V^2+ \mu_2^*{V^*}^2\right)\ . \la{hp}
\eeqa
 The equations of motion for the 2-d metric components determine $P_t$ 
(the momentum conjugate to the $AdS_5$ time coordinate $t$) 
as a 
function of the other variables,  
and also imply that  $D_1\alpha=0$. Setting  $D_1\alpha=0$ in \rf{hp} 
 we obtain
\beqa
\LL &=& -P_t + P_{\alpha}\dot{\alpha } 
 - i P_{\alpha} V^*   \dot{V}- \frac{1}{2\sqrt{-g}g^{00}} \bigg(-P_t^2 
   +P_{\alpha}^2 + |D_1 V|^2 \nonumber \\
  && + \ \half\left[e^{2i\alpha} (D_1V)^2 + e^{-2i\alpha}(D_1V^*)^2\right]\bigg)  
- \mu_1 (VV^* -1) -\half \left(\mu_2V^2+c.c.\right)\ .  \la{Lagr} 
\eeqa
 We  see that $P_\alpha$ is indeed the momentum conjugate to $\alpha$.
 Also,  the momentum conjugate to $V_n$ is then 
$$\PV{n} = -i P_\alpha V^*_n \ . $$
 Thus, 
 $V$ and  $V^*$ are not canonically conjugate  variables since,  
in general,  $P_{\alpha}$ is not constant. For later reference  let us list also 
the non-vanishing Poisson brackets 
\beqa
{[}\alpha(\sigma,\tau),P_{\alpha}(\sigma',\tau)] &=& \delta(\sigma-\sigma') \ , \nonumber \\
{[}\alpha(\sigma,\tau),V^*_n(\sigma',\tau)] &=& 
{[}\alpha(\sigma,\tau),\frac{i\PV{n}(\sigma',\tau)}{P_{\alpha}(\sigma',\tau)}] = 
-\frac{V^*_n(\sigma,\tau)}{P_{\alpha}(\sigma,\tau)}\delta(\sigma-\sigma')\ , \la{PB}\\ 
{[}V_m(\sigma,\tau),V^*_n(\sigma',\tau)] &=& [V_m(\sigma,\tau),\frac{i\PV{n}(\sigma',\tau)}{P_{\alpha}(\sigma',\tau)}] 
= \frac{i}{P_{\alpha}(\sigma,\tau)} \delta_{mn}\delta(\sigma-\sigma') \ .\nonumber
\eeqa
 Let us recall that we wanted to associate $P_\a$ with a large conserved momentum 
which characterizes the solutions
 but, due to the $\a$ dependence of the Lagrangian (\ref{Lagr}),
 the equation of motion for $\a$ implies 
that $P_\a$ is not conserved.
To identify the relevant  conserved quantity let us do a coordinate transformation 
that preserves
(up to a total time derivative) 
the form of the terms $P_{\alpha} (\dot{\alpha}  - i V^*\dot{V} )$ 
in \rf{Lagr} but  at the same time
eliminates all $\alpha$ dependence in the rest of the Lagrangian. 
The new $P_\a$ will be conserved
and can be used to classify the solutions. 
Unfortunately, it is not clear to us if this change 
of coordinates can be done in a closed manner
 but it can  certainly  be done order by order 
in an expansion in $1\ov P_{\alpha}$ as we explain below.

 Even perturbatively, finding this transformation 
directly is hopelessly lengthy if we do not resort to the method of 
canonical perturbation theory   which we review in  Appendix following \ci{LL}. 
The  infinitesimal canonical  transformation
is determined by a function $W(\bar{y})$ through 
\beq \la{delta}
\delta\bar{y} = \bar{y}_{_{\mbox{new}}} - \bar{y}_{\mbox{old}} = [\bar{y},W(\bar{y})] \ , 
\ \ \ \ \ \ \   \bar{y} \equiv  (\alpha,P_{\alpha},V,V^*)\ ,
\eeq
where we used the (non-canonical) Poisson brackets of eq.(\ref{PB}).\footnote{The fact
 that our  variables are 
non-canonical is not a problem for applying the  perturbation theory \ci{LJ}.}
We cannot use an arbitrary
function $W$ since we want, at the same time, 
to preserve the constraint $V^2=0, \ {V^*}^2=0$  but  generically 
$[V^2,W]\neq 0$. Instead of restricting the function $W$ 
we can define a different transformation using the Dirac 
bracket\footnote{This is equivalent to adding  to $\delta V$ a term 
proportional to $V^*$ which trivially preserves 
the terms linear in time derivatives and can be chosen such that $\delta(V^2)=0$.} 
since then the constraints 
commute with all functions. The definition 
of this bracket is
\beq
{[}f ,g]_{_D} = [f,g] -\frac{iP_{\alpha}}{4} [f,V^2][{V^*}^2,g] +  \frac{iP_{\alpha}}{4}[f,{V^*}^2][V^2,g] \ .
\eeq
Then using  \rf{PB} we find the  following non-vanishing brackets
\beqa
\left[\alpha(\sigma,\tau),P_{\alpha}(\sigma',\tau)\right]_{_D} &=& \delta(\sigma-\sigma') \ ,\nonumber\\
{[}\alpha(\sigma,\tau), V^*_n(\sigma',\tau) ]_{_D}
           &=& -\frac{ V^*_n (\sigma,\tau)}{P_{\alpha}(\sigma,\tau)} \delta(\sigma-\sigma') \ ,\la{DB}\\
{[}V_m(\sigma,\tau),V^*_n(\sigma',\tau)]_{_D} &=& \frac{i}{P_{\alpha}(\sigma,\tau)} \bigg[ \delta_{mn} 
                 - V^*_m(\sigma,\tau) V_n(\sigma,\tau)\bigg] \delta(\sigma-\sigma') \ .\nonumber
\eeqa
Note that these brackets are of order $1\ov P_{\alpha}$ in the sense that the right  hand side contains
one less power of $P_{\alpha}$ than the left  hand side.
 This means that, upon quantization
 (which would require also to include fermionic terms), the commutators will 
vanish in the limit $P_{\alpha}\rightarrow \infty$ which can then be understood as
 a classical 
limit.\foot{It also follows that in the quantum treatment $V^*$ should roughly be identified 
with $\frac{\delta}{\delta V}$ 
which  makes the structure of the 
Hamiltonian  very reminiscent of the form of the dilatation operator discussed in \ci{bks} 
in terms of the scalar  field $\Phi_i$ and $\breve{\Phi}_i$ defined as an operator that
 ``destroys'' a scalar field.}
 
We are still to satisfy the constraint $VV^*=1$, but 
this constraint is preserved if
\beq
{[} VV^*,W]_{_D} = \frac{1}{P_{\alpha}} \left(\frac{\delta W}{\delta \alpha} +iV^*\frac{\delta W}{\delta V} 
 -iV\frac{\delta W}{\delta V} \right)=0 \ ,
\label{Wgaugeinv}
\eeq
which expresses the fact that $W$ has to be gauge invariant, i.e. invariant under \rf{gaugeinv}.
 
Let us now  perform the  transformation \rf{delta} on the Lagrangian \rf{Lagr}.
 In principle,  we need a finite transformation,
but if we expand it  in powers of $1\ov P_{\alpha}$ then, at lowest order,  we only need
an infinitesimal transformation  generated by some $W=W_1$ 
(we explain this in more detail in  Appendix, see eqs. (\ref{Weq}) 
and (\ref{Hpeq})). To transform the Lagrangian we note that a generic function $f(\bar{y})$ transforms
in such a way that
\beq
f_{_{\mbox{new}}}(\bar{y}_{_{\mbox{new}}}) = f_{\mbox{old}}(\bar{y}_{\mbox{old}}) \ \ \Rightarrow \ \ 
f_{_{\mbox{new}}} = f_{\mbox{old}} + \frac{\partial f_{\mbox{old}}}{\partial \bar{y}_j}\, [W_1,\bar{y}_j]_{_D} = 
f_{\mbox{old}} + [W_1,f_{\mbox{old}}]_{_D} .
\eeq
Note the opposite  order of $W$ and $\bar{y}$ in the Dirac bracket 
as compared  to (\ref{delta}).
 Since $P_t$, the terms linear in derivatives in \rf{Lagr}, and the constraints, are all 
invariant under such transformations,  then the only terms which transform are
\beqa
\LL_0 = P_{\a}^2\ , \ \  \ \ \langle \LL_1\rangle = |D_1 V|^2\ , \ \ 
\ \ \{\LL_1\} = \half\left[e^{2i\alpha} (D_1V)^2 + e^{-2i\alpha}(D_1V^*)^2\right]\ ,  \la{dq}
\eeqa
which appear in the Lagrangian \rf{Lagr} 
multiplied by a common factor which is invariant.
Here $\langle ...\rangle$ denotes terms  averaged over $\a$ and therefore $\a$-independent, 
and
$\{...\}$ denotes terms which are oscillating in $\a$ and thus average to zero.

The leading term in the large $P_\a$ expansion  is $\LL_0$,   so
 if we want to cancel the $\alpha$ dependence at the 
lowest order, we need a function $W_1$ such that
\beq
[W_1,\LL_0]_{_D} = [W_1,P^2_{\alpha}]_{_D} = 2P_{\alpha} \frac{\delta W}{\delta\alpha} = -\{\LL_1\}=
     -\half \left[e^{2i\alpha}(D_1V)^2+e^{-2i\alpha}(D_1V^*)^2\right] \ .
\eeq
Integrating this  equation, we obtain $W_1$ 
\beq
W_1 =  \frac{i}{8} \int d\sigma\,\frac{1}{P_\a}
 \left(e^{2i\alpha}(D_1V)^2-e^{-2i\alpha}(D_1V^*)^2\right) \ ,
\label{W1e}
\eeq
which is (manifestly) gauge invariant and
 so preserves the constraint $VV^*=1$ as required. 
We have used that $D_1 \alpha=0$.
The transformed Lagrangian, written  in the new variables, 
 is the same as the original one written in the old 
variables, except that the term $\{\LL_1\}$ is no longer present. 
The end result of this  procedure is thus equivalent to averaging 
the Lagrangian over $\alpha$ as we did in the previous subsection. 

To go beyond the leading order  we use the results
described in  Appendix (see eq.(\ref{Hpeq})) and compute the second order term as 
\beq
\bar{\LL}_2 = \half\  \langle \ [W_1,\{\LL_1\}]_{_D}\  \rangle \ , 
\la{LL2}
\eeq
where, as before, $\langle...\rangle$ denotes averaging over $\a$.
This follows from the second order contribution of $W_1$ to the transformation. There is 
also a $W_2$ contribution to $W$ at this order but it  just cancels the $\a$ dependence in 
the second order terms,  as $W_1$ did for the first order ones. 
 Therefore, what we need to compute is 
\beqa
&&\bigg\langle\frac{i}{32}\left[\int d\sigma \frac{1}{P_{\alpha}}
\left(e^{2i\alpha}(D_1V)^2-e^{-2i\alpha}(D_1V^*)^2\right),\int d\sigma' 
\left( e^{2i\alpha}(D_1V)^2+e^{-2i\alpha}(D_1V^*)^2\right)\right]_{_D}\bigg\rangle  \nonumber\\
&&\ \ \ = -\frac{1}{4P_{\alpha}^2} \left(D_1^2VD_1^2V^*-(V^*D_1^2V)(VD_1^2V^*)-\frac{3}{2}(D_1V)^2(D_1V^*)^2\right)\ ,
\eeqa
where we used  that $D_1\alpha =0$ and anticipated that we are going to choose a gauge 
where $P_{\alpha}=$const by taking $D_1P_{\alpha}=\partial_1P_{\alpha}=0$.
Replacing in the Lagrangian we get
\beqa
\LL &=& -P_t + \dot{\alpha}P_{\alpha} - i P_{\alpha} V^*   \dot{V}- \frac{1}{2\sqrt{-g}g^{00}} \bigg[-P_t^2 
   +P_{\alpha}^2 + (D_1V)(D_1V)^*  \nonumber \\
  && -\frac{1}{4P_{\alpha}^2} \left(D_1^2VD_1^2V^*-(V^*D_1^2V)(VD_1^2V^*)-\frac{3}{2}(D_1V)^2(D_1V^*)^2\right)
 + \cO(\frac{1}{P_{\alpha}^3}) \bigg]  \nonumber \\
 &&- \mu_1 (VV^*-1) -\half \left(\mu_2V^2+c.c.\right)\ . 
\eeqa
Now we can choose a gauge $P_{\alpha}=\bL$=const as in \ci{krt} 
(here $\bL$ is a generalization of $\J$ in \rf{lama}, i.e. $L= \sql \bL$
as in \rf{vbvb}) 
 and substitute the expression for 
$P_t$ from the constraint following from variation over $g^{00}$, 
  or do the  T-duality transformation   $\a \to \tilde \a$  and fix 
the static gauge $ \tilde \a = \bL \s$ 
as discussed in  section 2. 
In both cases we get the following Lagrangian (cf. \rf{eaa}) 
\beqa
\LL &=& - i \bL V^*   \dot{V} - \sqrt{h} - \mu_1 (VV^*-1) -\half \left(\mu_2V^2+c.c.\right)\  , \\
h &=& \bL^2  + |D_1V|^2 
 -\frac{1}{4\bL^2} \left(D_1^2VD_1^2V^*-(V^*D_1^2V)(VD_1^2V^*)
-\frac{3}{2}(D_1V)^2(D_1V^*)^2\right) \nonumber \\
 && +\  \cO(\frac{1}{\bL^3}) \ .
\eeqa
The corresponding action takes the same form as \rf{vbvb} which generalizes \rf{eaa}.
Expanding for large $\bL \equiv  { 1 \ov \sqrt{\tilde \lambda}}$ 
is now a straightforward Taylor expansion of $\sqrt{h}$.
 Notice,  however, 
 that we can only
do the expansion up to order $\frac{1}{\bL^3}$ since we did not 
compute further terms in the canonical transformation.
The final result for the first two leading terms in the effective Lagrangian 
is thus (we omit a constant order $\bL$ term)
\beqa
\LL &=& \bL \bigg[- i  V^*   \dot{V}  - \frac{1}{2\bL^2} D_1V D_1V^* \nonumber \\
 && +\  \frac{1}{8\bL^4} \bigg( (|D_1V|^2)^2  + |D_1^2V|^2 
-|V^*D_1^2V|^2 -
    \frac{3}{2}|(D_1V)^2|^2\bigg) \bigg] \nonumber \\
 &&-\  \mu_1 (VV^*-1) -\half \left(\mu_2V^2+\mu_2^*{V^*}^2\right)\ . 
\la{LLL2}
\eeqa
In the special case of  $SU(2)$ sector of string states with  two angular momenta
 this Lagrangian can be seen to agree with the Lagrangian 
 found in \ci{krt} (and also with \rf{subb} in $SU(3)$ sector, 
upon using  leading-order  equations there to eliminate the
 time derivative terms). 
  One can also check that this Lagrangian does reproduce 
the large $\bL$ expansion of  the exact pulsating solution discussed in section 6.

 At this order the Hamiltonian follows from (\ref{LLL2}) as (we now include the constant term $\bL$):
\beq
H = \bL +\int_0^{2\pi} \frac{d\sigma}{2\pi} \left\{\frac{1}{2\bL^2} |D_1V|^2  +\  \frac{1}{8\bL^4} \bigg( (|D_1V|^2)^2  + |D_1^2V|^2 
-|V^*D_1^2V|^2 - \frac{3}{2}|(D_1V)^2|^2\bigg) \bigg]\right\}
\eeq
If we find a solution of the equations of motion for this Hamiltonian, or equivalently the Lagrangian (\ref{LLL2}),
we can obtain the time evolution of $\alpha$ as
\beq 
\alpha = \frac{\partial H}{\partial \bL} t = t + \omega_{\alpha}
 t\ , 
\eeq
where we denoted the (small) correction as $\omega_{\alpha}= \partial(H-\bL)/\partial \bL$. If we now want to know the profile of the 
string $X_m(\sigma,t)$ we should first undo the canonical transformation (\ref{delta}) and then use (\ref{xp}).
This results in 
\beq
X_m = \frac{1}{\sqrt{2}} \left\{ e^{it+i\omega_{\alpha} t} \left[V_m-\frac{i}{\bL} 
\left(\delta_{mn}-V^*_m V_n-V_mV^*_n\right)\frac{\delta W_1}
{\delta V^*_n}  \right] + \mbox{c.c.}\right\}\ , 
\label{Xf}
\eeq
where we can further use that (see (\ref{W1e})) 
\beq
\frac{\delta W_1}{\delta V^*_n} =
 \frac{i}{4\bL} e^{-2i\alpha} D_1^2 V^*_n\ . 
\eeq
As follows from the discussion in 
the next section, it is natural to assume that the transformed variables are identified 
with similar variables on the spin chain side. 
If that is so,\footnote{We are grateful to 
  A. Mikhailov for a question on this issue.}
(\ref{Xf}) can be understood as a change of variables
 from the spin chain variables $V_m(\sigma,t)$ 
to the string configuration $X_m(\sigma,t)$. At leading order we see from 
(\ref{Xf}) that the transformation is simply (\ref{xp}), the first 
correction being of order $\bL^{-2}$. 

As a final comment we note that to obtain 
the expression (\ref{Xf}) we expanded the exponent in 
\beq
e^{i\alpha-i[\alpha,W_1]} \simeq e^{i\alpha} (1-i[\alpha,W_1]+\ldots)
\eeq
and used the property (\ref{Wgaugeinv}). 
This expansion is valid at this order,  but, as discussed before, further expanding 
\beq
e^{it + i\omega_{\alpha} t } \simeq e^{it} (1 + i\omega_{\alpha} t +\ldots)
\eeq 
would be incorrect since secular terms, \ie\ terms linear in $t$, would appear.  
 
\section{Field theory side: effective action from $SO(6)$  spin chain \la{FT}} 

Our general aim is to compare the energies  of strings moving fast in the $S^5$ part of  
\adss  with the anomalous dimensions of the corresponding 
operators in the scalar sector of \N{4} SYM theory. The operators we are interested in
can be written as  
\beq \la{ops}
\cO = C_{m_1\ldots m_L} \tr \left(X^{m_1}\ldots X^{m_L} \right) \ , 
\eeq
where $X_m$ ($m=1,\ldots,6$) are real scalar fields in adjoint representation of 
$SU(N)$ and the coefficients $C_{m_1\ldots m_L}$ are complex in general.
 If we just wanted operators that have a well defined
anomalous dimension, namely eigenvectors of the dilatation operator,
then, since $\cO^\dagger$ has the same conformal dimension as $\cO$,  
we could always choose the coefficients $C_{m_1\ldots m_L}$ to be 
real. However, this is not possible if, for example, we want to find operators
of given R-charges such as,  \eg,  tr$ X^J$ where $X=X_1+iX_2$. 
 Also, as described in \ci{kru,krt,ST}, in our case we are not 
looking for operators which are eigenstates of the dilatation operator but 
for ``coherent''  operators that behave under the renormalization group evolution as 
the classical string does under the  time evolution.
 This can be understood more
easily if we do a conformal mapping of the field theory to $R\times S^3$. In that 
case the operators we are looking for create
 coherent 
states which are time dependent in 
exactly the same way as the classical string solutions are. So the conclusion is that,
for our purpose, we cannot restrict the coefficients $C_{m_1\ldots m_L}$ to be real.

To proceed, consider the 1-loop dilatation operator acting on the scalar operators \rf{ops} 
which is equal (in the large-N limit) to  \cite{mz1} 
\beq
H_{m_1\cdots m_L,\, n_1\ldots n_L} = \frac{\lambda}{(4\pi)^2} 
  \sum_{l=1}^{L} \left(\delta_{m_lm_{l+1}}\delta^{n_ln_{l+1}}+
2\delta^{n_l}_{m_l}\delta^{n_{l+1}}_{m_{l+1}}
-2\delta^{n_{l+1}}_{m_l}\delta^{n_{l}}_{m_{l+1}}\right)\ .
\label{Hamil}
\eeq
It  has a nice interpretation  \ci{mz1} as an integrable Hamiltonian 
acting on an $SO(6)$ spin chain.
Its eigenvalues can be computed using Bethe ansatz techniques \cite{mz1}.
 For the purpose of
comparison with string theory it is better, however, 
 to use an effective action
 approach  which allows one  to describe the 
low energy states in a universal way. 
This action can be directly compared  with  (a limit of) the 
 semiclassical \adss\  string  action. Therefore,  it is appropriate to say that,
in this method, we compare semiclassical solutions on one side with 
semiclassical solutions on the other.

To obtain this  effective action description one can follow 
\ci{kru,krt,ST} and use the method of coherent states    (see, e.g.,  \ci{zhang}).
This method produces a path integral representation
for the propagator and therefore one gets a classical action 
for the system. As an approximation one can just consider classical
solutions which is justified since through a rescaling of the coordinates   
one can see that the inverse of the large momentum acts as the effective Planck 
constant of the system. A similar idea we discussed in the previous section 
regarding the commutators (\ref{DB}). Instead of following this path 
we can reach the same result by an alternative method discussed in \ci{krt}
which is also more suitable for computing corrections to the leading order
result (even though we are not going to do this  in the present  paper).

 The first step is to use a factorized ansatz for the operator \rf{ops}
where the matrix $C_{m_1\ldots m_L}$  is given by the product of 
$L$  6-vectors  $v_1,\ldots,v_L$
\beq\label{factorized}
\cO =  \tr ( \prod^L_{l=1}  v_{l m} X^{m} )
 \ , \ \ \ \ {\rm i.e.} \ \ \ \ 
C_{m_1\ldots m_L} = v_{1 m_1} \ldots v_{L m_L} \ ,\  \ {\rm 
or } \ \ \ \  C = \bigotimes_{l=1}^L v_l \ .
\eeq 
The case of BPS operators  for which $C$ should be totally symmetric and traceless
corresponds to all $v_l$ being equal to the same vector 
 $v$ with $v^2=0$.  

Once more, it proves useful to think of the theory in $R\times S^3$ where we deal
with states instead of operators. Each scalar particle can be in $6$ possible states $\ket{m},\
 m=1,\ldots, 6$ (since the fields are real and therefore the 
antiparticles are the same as the particles). A generic state of a 
particle is parameterized by six complex numbers up to a normalization condition 
and an overall irrelevant phase. 
The operator with coefficient matrix 
$C$ corresponds to a 
composite  state of $L$ particles each in a state given
 by $v_m\ket{m}$. The normalization condition implies $v v^*=1$ and the 
overall irrelevant phase translates into the gauge invariance 
$v \to e^{i \b} v $ as in (\ref{gaugeinv}). 
 Therefore, the vectors $v $ effectively live in the 10-dimensional space $CP^5$. 

The Schr\"odinger  equation for the wave function of a state associated to 
$C$ follows from minimizing the action
\beq
S = -\int dt \left(  i 
C^*_{m_1\ldots m_L}
 \frac{d}{dt} C_{m_1\ldots m_L} 
  +
C^*_{m_1\ldots m_L} H_{m_1\cdots m_L,\,
 n_1\ldots n_L} C_{n_1\ldots n_L} \right) \ . 
\eeq 
In general, variations of $C$ with $t$ may be interpreted as a 
RG evolution of the ``coupling constants'' corresponding to the scalar operators \rf{ops} 
(with the canonical dimension factor extracted). 

Instead of minimizing the action $S$ in the full space of tensors $C$ we shall consider
its  reduction  to the subsector given by the factorized 
ansatz
 (\ref{factorized}).
Then we find (suppressing 6-vector indices on $L$ vectors 
$v_l$ each  satisfying   $v^*_l v_l=1$)\foot{The 
same expression is found   by starting with the form of 
the dilatation operator \rf{Hamil} written in terms of the $SO(6)$ generators
  in the 
vector representation $M^{ab}_{mn}= 
\delta^{a}_m \delta^{b}_n -  \delta^{b}_m \delta^{a}_n$ \ci{mz1}
$H= { \l \ov (4 \pi)^2}\sum_{l=1}^{L} H_{l,l+1}  ,\ \ 
H_{l,l+1} =
M_l^{mn}M_{l+1}^{mn}-\frac{1}{16}(M_l^{mn}M_{l+1}^{mn})^2+\frac{9}{4}$
and replacing $M^{mn}_l$ by $ <v|M^{mn}_l|v>$ or $ v^m_l v^{*n}_l -  v^n_l v^{*m}_l$
with $v^*_l v_l=1$.}
\beqa \la{acta}
S = - \int dt \sum_{l=1}^{L} \left\{ i \bv_l \frac{d}{dt} v_l + 
   { \l \ov (4\pi)^2} 
\bigg[(\bv_l\bv_{l+1})(v_lv_{l+1}) + 2  - 2 (\bv_lv_{l+1})(v_l\bv_{l+1})\bigg]
 \right\} \ . 
\eeqa
which is gauge invariant under $v_l\rightarrow e^{i\beta_l(t)} v_l$. 
As expected \ci{mz1}, the Hamiltonian (the term in the square brackets)
 vanishes  for the 
BPS  case when $v_l$ does not depend on $l$, i.e.   $v_l=v$, and  $v^2=0$.
More generally, if we assume that $v_l$ is changing slowly with $l$ (\ie\ $v_l\simeq v_{l+1}$),  
then we find that \rf{acta} contains a  potential term 
$(\bv_l\bv_{l})(v_lv_{l})$ coming from the first ``trace'' structure in \rf{Hamil}.
This  term will lead  to large (order $\l L$ \ci{mz1}) 
shifts of anomalous dimensions, 
invalidating a low-energy  expansion, i.e. prohibiting one from taking the  
continuum limit  
\be \la{cont}
L\to \infty\ , \  \ \ \ \ \ \ \ \tl={\l\ov L^2}={\rm fixed} \ , \ee  
 and thus from establishing a  correspondence with string theory 
along the lines of \ci{kru,krt,ST}.\foot{Note that the same expression 
\rf{acta} in the case of real vectors $v_l$ was found in \ci{ST} as 
corresponding  to the second $SO(6)/SO(5)$ choice of the $SO(6)$ coherent state 
with  the non-BPS ``ground state''  $v=(0,0,0,0,0,1)$. 
In this case $\bv_l=v_l $ so that the first  (WZ-type) term in \rf{acta} vanishes 
and the Lagrangian in \rf{acta} becomes simply ${\l\ov (4 \pi)^2}
 \sum_{l=1}^{L}
 [2-(v_l v_{l+1})^2]$. Then 
the potential term is  constant and proportional  to $\l L$. In this case  one 
cannot consistently define the continuum limit \rf{cont}. 
This real case is a consistent truncation of the general complex case 
we consider here provided ${d \ov dt} v_l=0$
(this does not imply the vanishing of the anomalous dimension because 
$v_l$ are constrained by $v^2_l=1$, i.e. there is an additional
 Lagrange multiplier term 
in the equation for $v_l$).}

To get the low energy solutions
when  variations of $v_l$ from site to site are small we are thus 
to impose the restriction
\be \la{rest}
v^2_l=0 \ , \ \ \ \ \ \ \ \ \ \
l=1,\ldots,L \ee
 which minimizes the potential energy coming 
from the first term in the Hamiltonian (\ref{Hamil}).
Note that, like $v^*_l v_l=1$,  this condition  is to be imposed  in a ``strong''
 sense, i.e. with a Lagrange multiplier: 
one can check that if imposed at $t=0$ this condition 
 is not preserved by the time  evolution 
implied by the equation following from \rf{acta}. 
This does not, however, imply a problem of principle for  our present goal
of comparison with semiclassical states in string theory: 
mixing between states with $v^2_l=0$ and with $v^2_l\not=0$ will be suppressed 
in the continuum limit \rf{cont} we are interested in
 (as already mentioned above,
 states with $v^2_l\not=0$  will have large
  anomalous dimensions in this limit, and the same will apply 
to off-diagonal elements 
of the Hamiltonian, i.e. of  the anomalous dimension matrix).

The condition \rf{rest}
implies that the operator at {\it each} site is 
invariant under half of the  supercharges of the $\cal N$=4 superalgebra. 
That can be seen from the supersymmetry variation of the 
operator $v_m X^m $:
\beq
 \delta_\epsilon (v_m  X^m )= \frac{i}{2} \bar{\epsilon} v_m \Gamma^m \psi \ , 
\eeq
where the index $m$ is summed from $1$ to $6$. If $v^2=0$, 
the matrix $v_m\Gamma^m$ satisfies $(v_m\Gamma^m)^2=0$.  
Then  half of its
 eigenvalues are equal to zero,  meaning that the operator 
$v_m X^m $ is invariant under the supersymmetry variations 
associated with the null eigenvalues.
We may thus call \rf{rest}  a ``{\it local} BPS'' condition
(see also \ci{kru,mikk,mikkk})
 since the preserved combinations of supercharges 
in general are different for each $v_l$ and therefore the 
complete operator corresponding to  $C=\bigotimes v_l$ is not BPS.
 ``Local'' should be
 understood in the sense of the spin chain,  or,  
equivalently,  the string world-sheet direction.\foot{This 
generalizes the argument implicit in \ci{kru};
equivalent proposal was made in \ci{mikk}. This is related to but different 
from the ``nearly BPS'' operators  discussed in \ci{mateos} 
(which,  by definition,   were those which become BPS in the limit $\l\to 0$).}

As already mentioned above, such states correspond 
to $SO(6)/[SO(4)\times SO(2)]$ coherent  states  of $SO(6)$ 
considered in \ci{ST,mikkk} which are obtained
(at each site $l$)  by applying an  $SO(6)$ 
rotation parametrized by an  element of the coset $G_{2,6}$ 
to the vacuum state $v=(0,0,0,0,1,i)$ satisfying $v^2=0$
(note that the condition \rf{rest} is invariant under $SO(6)$ rotations). 
The corresponding coherent state action \ci{ST} is then indeed equivalent 
to \rf{acta} with the corresponding constraints added,  i.e. 
\beqa \la{acti}
S &=& - \int dt \sum_{l=1}^{L} \left\{ i \bv_l \frac{d}{dt} v_l  +
   { \l \ov (4\pi)^2} 
\left[ (\bv_l\bv_{l+1})(v_lv_{l+1}) + 2  - 2 (\bv_lv_{l+1})(v_l\bv_{l+1})\right]
 \right. \nonumber \\ 
&& \left. +\ \mu_1  (\bv_l v_l-1) +  
\frac{1}{2} (\mu_2  v_l^2 + c.c) \right\}
 \ . 
\la{aca}
\eeqa
If we now consider the case when 
 $v_l$ are slowly varying then we
can take the continuum limit \rf{cont}  as in \ci{kru,krt,ST}
by  introducing the 2-d field $v_m(t,\s)$ with 
$v_{ml} (t)= v_m(t, {2\pi l\ov L})$. Then    
\beq \la{joj}
S = - L \int dt \inti  
\left\{ i\bv \frac{\partial v}{\partial t}  +  {1 \ov 2} \tl  (D_1v)(D_1v)^* 
 + \mu_1 (\bv v-1) +  
\frac{1}{2} (\mu_2 v^2 + c.c) \right\}
 \ , \eeq
where $\tl \equiv { \l \ov L^2}= { 1\ov \bL^2}$ and $D_1v = \partial_\sigma v - (v^*\partial_\sigma v)v$
as in (\ref{covdev}).
 Note that  all higher derivative terms are suppressed in the limit \rf{cont}
by powers of $1\ov L$ and the potential term 
is absent due to the condition $v^2=0$. 
Thus \rf{joj} becomes 
  equivalent to the $G_{2,6}$  ``Landau-Lifshitz'' sigma model \rf{hyy}
 obtained from the string theory action $I$ \rf{vbvb}
after we rename  $v_m \to V_m$ 
  and rescale   the 
time coordinate  by $\tl^{-1}$ as in \rf{resc} (and appropriately rescale  
the  Lagrange multipliers). This action has gauge invariance $v_m \to e^{i \beta} v_m$, 
where $\beta$ is an arbitrary function of $\sigma$ and $t$. 

 The equations of motion following from \rf{joj} are
\be \la{vev}
i \dot{v}_m  -  {1 \ov 2} \tl  \left[  v''_m   - 2 (\bv v') v'_m  \right] 
 + \mu_1   v_m +  \mu_2 ^*  v_m^* =0 \ , \ee
where $v^*_m v_m =1, \ v^2_m=0, \ v^*{}^2=0$ and dot and prime
stand for derivatives over $t$ and $\sigma$. 
Multiplying this by $v^*_m$ and combing with complex-conjugate 
equation we learn that $(v^*_m v'_m)'=0$, i.e. $v^*_m v'_m= f(t)$, 
and also find the expression for $\mu_1$. Multiplying \rf{vev}
by $v_m$ we get $\mu^*_2 = - {1 \ov 2} \tl    v'_m v'_m$.  

 Besides the equations of motion, we need to add an additional condition.
The presence of the trace in (\ref{ops}) implies that 
we have to consider only spin chain states that are
invariant under translations in $l$ or in $\sigma$. 
Quantum
mechanically, this  means that the momentum in the 
direction $\sigma$ should vanish: $P_\sigma=0$.
 In the low energy effective action description we use here, 
directly 
imposing the translation invariance in $\sigma$  would be 
 too restrictive
(this  would leave us only with  constant fields), 
but we should still impose the global constraint 
$P_\sigma=0$ since the 
conserved momenta should agree between the semiclassical (coherent) 
and exact quantum states.
 From (\ref{joj})
we can compute the translational 
Noether  charge $P_\sigma$ with the result:
\beq
P_{\sigma} = \int_0^{2\pi} \frac{d\sigma}{2\pi} T_{01} = 
        L \int_0^{2\pi} \frac{d\sigma}{2\pi} \ iv^*_m  \partial_\sigma v_m 
 =0 \ .
\la{P=0b}
\eeq
where $T_{01} = - {\del  {\cal L} \ov \del  \dot v}  v'$. 
This should be viewed as  a condition on the solutions $v_m(t,\s)$. This condition
agrees precisely with 
(\ref{P=0}) that was obtained on the string side, after we identify  
$v_m$   with $V_m$ there.

In the following section we shall find particular solutions of equation 
\rf{vev} (and (\ref{P=0b}))
  and show that $S$ reproduces the one
loop anomalous dimensions obtained previously for rotating and pulsating 
strings using the Bethe ansatz \ci{bmsz,Min2,Min3}.
 We can thus conjecture
that, similarly to the previously discussed  pure-rotation 
$SU(2)$   and $SU(3)$  cases,  here 
one has an exact relation 
 between solitons of this ``Landau-Lifshitz'' sigma model \rf{joj}
and the corresponding  set of  Bethe ansatz eigenstates, 
generalizing the one  found in the $SU(2)$ case in  \ci{kmmz}.

Let us make an important comment on the interpretation of the operators 
associated, according to \rf{factorized},
 to the solutions of \rf{vev} and the corresponding string solutions. 
In the continuum limit we may write  the operator \rf{factorized}
 corresponding to the solution $v(t,\s)$ as 
\be \la{opa}
\cO =  \tr ( \prod_\s  \vv(t,\s) )  \ ,\  \ \ \ \ \ \ \ \
\vv \equiv  v_{m}(t,\s)  X^{m} \ .
\ee
As already mentioned above, such
a  ``coherent'' operator representing a coherent state of the spin chain 
will not in general be an eigen-operator of the dilatation operator or 
the Hamiltonian \rf{Hamil} (in particular, 
it may have a non-trivial dependence on $t$).
Indeed, as in flat space, where  a string state with a given  (large)  energy can be 
realized either as a pure  Fock eigenstate 
or as a coherent superposition of Fock space states, 
here the semiclassical string states 
represented by classical string solutions 
should be dual to coherent spin chain states or coherent operators, 
which are different from the exact eigenstates of the dilatation operator 
but which should lead to the same energy or anomalous dimension  expressions. 
Once again, the SYM operator naturally associated to a semiclassical string solution 
should be a locally BPS coherent operator \rf{opa}.
The $t$-dependence of the string solution thus translates into the RG scale 
dependence of $\cO$, while the $\s$-dependence describes the ordering of the factors under 
the trace.\foot{It should be  noted also  that not all ``long'' SYM operators 
(with large canonical dimension) 
are represented by the above locally BPS  coherent operators.
For example, BMN operators (e.g. $\sum c_k$ tr $( X^{J-k} Y X^k Y)$) 
are   dual to small fluctuations 
near point-like rotating string which are Fock-space states 
and which  are indeed exact eigenstates of the dilatation operator. 
Still, their energies can be reproduced by considering linearized solutions 
of the coherent state LL equation (i.e. by small fluctuations 
of unit vector $\vec n$ near $(0,0,1)$, see \ci{kru,krt}). The corresponding 
operators are then linearizations of the coherent operators 
corresponding to generic solution of the LL equation.
Note also that in contrast to ``good'' coherent operators dual to 
classical string states depending on (several) semiclassical parameters, 
the BMN states depend  only on one large quantum number $J$, and their 
anomalous 
dimensions scale as ${\l \ov L^2}, \ L=J,  $ and not as 
$  \l\ov L $ as in the case of ``true'' semiclassical operators.
}
At the same time, the Bethe ansatz approach \ci{bmsz,bfst,kmmz}
should be determining the  exact eigenvalues 
of the dilatation operator.
The reason why the two approaches happen to be in agreement 
is that in the limit \rf{cont}  we consider  the problem is 
essentially semiclassical, and because of  the integrability 
of the spin chain, its exact eigenvalues are not just 
well-approximated by the classical solutions
but are  actually exactly reproduced by them (just as in the harmonic oscillator or 
 flat space string theory case), i.e. 
the semiclassical coherent state  sigma model approach happens 
to be exact (related observations were made, e.g., in \ci{jev}).

\section{Particular leading-order solutions \la{part}}
\label{particularsolution}

In the previous section we have  found an agreement between the actions 
describing a string moving fast in the $S^5$ part of \adss and the action describing 
the continuum limit of the spin chain determining the 1-loop 
anomalous dimensions. 
 In this section we shall consider particular solutions of the equation \rf{vev}
\be \la{vev2}
i \dot{v}_m  =  {1 \ov 2} \left[  v''_m   - 2 (\bv v') v'_m  \right] 
-  \mu_1   v_m -  \mu_2^*  v_m^*  \ , \ee
which follows from the Lagrangian (\ref{hyy}) after replacing $V\rightarrow v$,  or, 
equivalently, from (\ref{joj}) after rescaling time by $\tl$ (see eq.(\ref{vev})). 
Such particular solutions include all rotating and pulsating string
configurations 
that have been discussed  in the literature and whose motion is only 
in the $S^5$.\foot{Pulsating solutions 
 in $AdS_5$ were considered in \cite{Min1,Smed}.}
The  description  of the   pulsating solution as 
a particular solution of the leading-order effective action 
 is new; as a result, we are able  
to explicitly identify the coherent SYM operators corresponding 
to pulsating solutions
(the form of the Bethe ansatz eigenstates 
associated to the same energy eigenvalues is implicitly  contained   
 in the previous work of  \cite{bmsz,Min2,Min3}).

\subsection{Rotating strings: $SU(3)$ and $SU(2)$ subsectors}

For generic rotating string solutions 
 the agreement between string theory and spin chain 
 at the level of the leading-order 
effective actions was already demonstrated  in 
\cite{kru,krt,ST,lopez}. 
Here we just show 
how this case  is included in our general 
treatment. To this end  we shall use the 
following ansatz for $v$ in \rf{joj},\rf{opa},\rf{vev2} 
\be 
v_m= \frac{1}{\sqrt{2}}
(U_1,\,iU_1,\,U_2,\,iU_2,\,U_3,\,iU_3) ,\ \ \ U^*_i U_i=1\  \Rightarrow  \ 
v_m v_m =0  , \ \ v^*_mv_m=1 
  \ . \ee 
This is in agreement with the expectations
 for a rotating string since $v$  can be written
also as 
\beqa \la{kopi}
v &=&   \frac{1}{\sqrt{2}}\left[U_1\, X + U_2\, Y + U_3\, Z \right] \ , 
\eeqa
where, as before, $X=X_1+iX_2$, $Y=X_3+iX_4$, $Z=X_5+iX_6$, and 
$(X_k)_m=\delta_{km},\  k=1,\ldots,6$ form a   basis in $R^6$.
Equivalently, this determines the form of v in \rf{opa}.
This ansatz is easily seen to satisfy the equations of motion 
\rf{vev2} if the complex 3-vector $U_i$, $i=1,2,3$ satisfies the equations of motion for the Lagrangian
\beq
\LL = - i U^*_i \dot{U}_i - \ha   \left[ U^*{}'_i U'_i + (U^*_i U'_i)^2\right]
 + \Lambda (U^*_i U_i-1)  \ ,
\eeq
which is just the Lagrangian in \rf{joj} after the substitution of \rf{kopi}.
This is the same  $CP^2$  Lagrangian  \rf{accv}  for leading-order 
 3-spin rotating solutions \ci{ST}  which 
 we  rederived  (and generalized to all orders)   in section 2.
The $SU(2)$ case of $CP^1$ model describing 2-spin solutions \cite{kru} is obtained
by setting  $U_3=0$. In that case a convenient parametrization is
\beq
U_1=\sin\frac{\theta}{2}\ e^{\frac{i}{2}(\phi-\varphi)}, 
\ \ \ \ \ U_2 = \cos\frac{\theta}{2}\ e^{\frac{i}{2}(\phi+\varphi)} \ .
\eeq
If we write the Lagrangian for these fields, $\phi$ disappears due to gauge invariance.  
The Lagrangian becomes
\beq
\LL = \half \cos\theta\  \dot{\varphi} - \frac{1}{8} ({\theta'}^2 + \sin^2\theta\  {\varphi'}^2) \ .
\eeq
It is an interesting exercise to evaluate the next correction 
by substituting  this ansatz into  (\ref{LLL2}) 
and thus reproduce the result of \ci{krt}.


\subsection{Pulsating strings: $SO(3)$ subsector}


To analyze the case of pulsating strings let us first discuss 
which operators should be dual to them. In the 
case \ci{Min1} where the string is pulsating  (with 
no rotation) we can consider the motion to be restricted to $S^2 \subset S^5$, 
i.e. to occur in the 
coordinates $X_1$, $X_2$, $X_3$, with $X_1^2+X_2^2+X_3^2=1$. 
Using the standard polar coordinates $(\theta,\phi)$ on the 
sphere, the solution is of the form 
$\theta=\theta(\tau)$,  $\phi=m\sigma$,  with 
$m$ being an  integer. 
While the exact 
 form of the function $\theta(\tau)$ 
is obtained  in section \ref{puls}, here we will  only need to know that it 
is approximately given by $\theta\simeq t$, so that 
$\theta$  plays the role of the fast coordinate.
 Namely,  
each piece (or bit) of the closed pulsating string moves along  a maximal radius 
 circle of constant longitude $\phi$ at almost the speed of light.
{}From the generic correspondence we discussed in the previous sections we expect
 that each piece of the string maps into the scalar operator \rf{opa} with $\vv$ 
of the
 following form 
\beq
\vv(\sigma) = \frac{1}{\sqrt{2}} \left[ \cos\phi\  X_1 
+ \sin\phi\  X_2 +  i X_3 \right] \ , \ \ \ \ \ \ \phi= m \s \ , 
\eeq
i.e. the corresponding 6-vector  is $v_m=\frac{1}{\sqrt{2}}(\cos \phi, \sin \phi, i, 0,0, 0)$
which satisfies $v^*v=1, \ v^2=0$.

We can check that this  is a solution of the reduced action by considering 
the special case  of  \rf{hyy} or, equivalently, 
  \rf{joj}  when the motion occurs only in 3 out of 6 directions, i.e. within 
 $S^2$. 
 In general, writing \rf{joj} in terms of 6 real coordinates 
\be \la{real}
v_m \equiv { 1 \ov \sqrt 2} ( a_m + i b_m) \ , \ \ \ \ \ \ \ 
a^2_m=1 \ , \ \ \ b^2_m=1\ , \ \ \ a_m b_m =0 \ , \ee 
we get (renaming the Lagrange multipliers) 
\be \la{trunc}
\LL  = a_m \dot b_m  - { 1 \ov 4}   \bigg[ a'^2_m + b'^2_m  - 2 (a_m b'_m)^2 \bigg]
+ \Lambda ( a^2_m-1) +\tilde  \Lambda (b^2_m -1) + \ \mu\  a_m b_m \ . \ee 
This Lagrangian still has  gauge invariance associated with an  arbitrary phase transformation
of $v$, allowing one to impose one real gauge condition. That leaves us with 
$2\times 6 - 3 -1 =8$ real independent functions.

Let us  now  restrict to the case when 
$m$ runs only 1,2,3, i.e. the corresponding string motions happen in $S^2$
part of $S^5$. Then  we should have $2\times 3 - 3 -1 =2$  independent 
real variables. We can fix the gauge invariance by the condition $b_3=0$;
then the unit 3-vector $b=(b_1,b_2,0)$ orthogonal to the unit vector 
$a=(\sin \vartheta \ \cos \phi, \sin \vartheta \ \sin \phi, \cos \vartheta)$
can be chosen as  $b=(\sin \phi,- \cos \phi,0)$. Then \rf{trunc} is found to be 
 \be \la{unc}
\LL = \sin\vartheta\    \dot \phi 
  - { 1 \ov 4}  ( \vartheta'^2 + \cos^2\vartheta\ \phi'^2) \ ,  \ee 
which is formally equivalent to the LL sigma model \ci{kru,krt} 
found in the $SU(2)$ rotating sector where strings are moving in $S^3$ 
part of $S^5$. There, one of the three angles of $S^3$ was a fast variable, 
and eliminating it, we were left with two dynamical variables.
Here, we have two coordinates of $S^2$ to start with;    
eliminating one fast coordinate we are left with a ``slow'' coordinate, 
but  we are still  to keep 
{\it one more}   degree of freedom which may be interpreted as a  momentum
conjugate to the ``slow'' coordinate.

A special solution is $\vartheta={\pi \ov 2}, \ \phi= m \sigma$ which 
corresponds to the pulsating solution  we discussed
above. Taking into account that, with the Lagrangian (\ref{unc}), 
the action is given by  (\ref{joj}), i.e. 
\beq
S = L \int dt \int \frac{d\sigma}{2\pi}\,\  \LL
\eeq
where $t$ is related to the time  in \rf{unc} by the rescaling \rf{resc} with 
  $\tl=\frac{\lambda}{L^2}$,  we obtain the energy of the pulsating  solution as
\beq
E = \frac{1}{4} \tl L\,  \cos^2\vartheta\  \phi'^2 = \frac{\lambda }{4L} m^2  \ , 
\eeq
which is in agreement with the result
 of \ci{Min1} (see also eq.(\ref{E(J,L)})  for  $J=0$.)


\subsection{Pulsating and rotating strings: $SO(4)$ subsector}


The case of a pulsating {\it and}   rotating string \ci{Min2} is slightly more involved. 
Here the motion occurs in $S^3\subset S^5$ parametrized by 
 $X_1,\ldots,X_4, \ X^2=1$. In the standard parametrization
\beq
X_1 = \sin\theta \cos\phi_1,\ \ \ X_2 = \sin\theta \sin\phi_1,
\ \ \ X_3 = \cos\theta\cos\phi_2,\ \ \ X_4 = \cos\theta\sin \phi_2,
\eeq
 the metric on  $S^3$ is 
\beq\la{sss}
ds^2 = d\theta^2 + \sin^2\theta\ d\phi_1^2 + \cos^2\theta \  d\phi_2^2\ , 
\eeq
and the string  ansatz is then 
\beq \la{ana}
\theta=\theta(\tau)\ , \ \ \  \phi_1=\phi_1(\tau),\ \ \ \ \phi_2 = m\sigma \ . 
\eeq
The closed string moves  approximately  at the speed of light along 
a tilted maximal  circle in the sphere $S^2$ 
parametrized by $(\theta,\phi_1)$. Let us denote the  angle
 between the normal to the circle and the $\theta={\pi\ov 2}$ plane
 as $\galpha$. In terms of conserved quantities it is given by
\beq\la{sa}
\sin\galpha = \frac{J}{L} \ ,
\eeq
where $J$ is the  momentum conjugate to $\phi_1$ and $L$ is 
defined through $L = J  + I_{\theta}$,
where $I_{\theta}$ is the action integral corresponding to the coordinate 
$\theta$ (see next section for more details). 

Using the same arguments as above, namely associating  each point of the string 
moving along a maximum circle with the corresponding operator carrying the same
R-charge, we can identify this string configuration 
with an operator as in (\ref{opa}) where $\vv$ is  given by:
\beq
\vv_{_{\mbox{leading order}}} = \frac{1}{\sqrt{2}}\left[i
 X_1 + \sin\galpha\  X_2 + \cos\galpha\ \cos\phi_2\  X_3 + \cos\galpha\ \sin\phi_2\
 X_4 \right]
\ , 
\label{leadingv}
\eeq 
where we have  emphasized that this identification 
is valid at the  leading order which 
is equivalent to considering 
 $\phi_2$ as being  constant (\ie\ $m=0$ in (\ref{ana})).  
For $\phi_2=m\sigma$, $m\neq 0$,  the 6-vector $v_m$  corresponding to  \rf{leadingv} 
is not a 
solution of the equation (\ref{vev2}).
 To get a consistent 
solution we use eq.(\ref{leadingv}) 
as a guide and propose to look  for $v_m$ as 
\beq
v = v_0(t) + w(\sigma) \ , 
\eeq
with $v_0$ being a $\s$-independent complex vector  having components along
 $X_1$ and $X_2$ only. The vector $w$ is real and has components along
 $X_3, X_4$. The constraints $v^2=0$
 and $v\bv=1$ imply
\beq
1-\bv_0 v_0 = w^2 = -v_0^2  \ ,
\la{cv0}
\eeq
 so that, in particular, $v_0^2$ and $|v_0|^2$ should be  time-independent.  
 Taking this into account, and using as a further assumption 
 that the Lagrange multipliers $\mu_1$ and $\mu_2$ are real 
constants we find that \rf{vev2} reduces to 
\beqa
i\dot{v}_0 =  -\mu_1 v_0 - \mu_2 v^*_0 \ ,\ \ \ \ \ \ \ \ \ \ 
 0= w'' - 2 (\mu_1+\mu_2) w \ .
\eeqa 
The equation for $v_0$ is easily solved. We get
\beq
v_0 = A e^{i\omega t} + B e^{-i\omega t} \ , 
\eeq
with $\omega=\sqrt{\mu_1^2-\mu_2^2}$ and $A$, $B$ are complex
 two component vectors that have to satisfy
\beq
A^2 = 0\ ,\ \ \ \ \ \ 
 B^* = -\frac{\mu_2}{\mu_1 +\omega} A\  , \ \ \  \ \ \ |A|^2 = \frac{(\mu_1+\omega)^2}{
(\mu_1+\omega+\mu_2)^2} \ .
\la{AB}
\eeq
For the other vector $w$ we can choose the expression 
 suggested by (\ref{leadingv}), i.e. 
$w = \frac{1}{\sqrt{2}} (0,0,\cos\galpha\cos m\sigma,\cos\galpha\sin m\sigma)$ 
which implies $\mu_1+\mu_2=-\half m^2$
and $w^2 =\half \cos^2\galpha$. Since $v_0^2 = -w^2$
 we can use (\ref{AB}) to find  that
\beq
\mu_1 = -\frac{m^2}{4} (1+\sin^2\galpha)\ ,\ \ \ \  \ \mu_2
=-\frac{m^2}{4}\cos^2\galpha\ , \ \ \ \ \   \omega=-\frac{m^2}{2}\sin\galpha\ .
\label{ABw}
\eeq
Choosing an arbitrary phase to obtain an operator similar to (\ref{leadingv}) we
 can then write the solution as
\beq
v = \frac{1}{\sqrt{2}} (-\sin\galpha\sin\omega t+i\cos\omega t,\ \,\sin\galpha\cos\omega 
t+i\sin\omega t,\,
   \cos\alpha\cos m\sigma,\,\ \cos\galpha\sin m\sigma) \ .
\la{vsol}
\eeq
 For this solution we can compute the angular momenta
$M_{mn} = \inti \M_{mn}$  using (\ref{MV}). 
The only non-vanishing densities
are:
\beqa
\M_{12}(\sigma) = L\sin\galpha\ , \ && \M_{13}(\sigma)+i \M_{23}(\sigma) =
 L\ e^{i \omega t}\cos m\sigma\cos\galpha\ ,  \\
&& \M_{14}(\sigma)+ i \M_{24}(\sigma) = L\ e^{i \omega t} \sin m\sigma\cos\galpha \ . 
\eeqa
Integrating over $\sigma$ we find that 
 the only non-vanishing angular momentum component is 
\beq
M_{12} = \int \frac{d\sigma}{2\pi} \M_{12}(\sigma) = L\sin\galpha = J \  , 
\eeq
in agreement with 
 the definition $\sin\galpha = {J\ov L}$ in  (\ref{sa}). 
While it was clear that  $J$ is the only non-vanishing angular
momentum, computing the densities $\M_{mn}(\sigma)$ is instructive since the
result agrees with
the expectations from the full string solution. 

The energy can be evaluated from (\ref{joj}) using that
\beq
v^*v' = 0\ ,\ \ \  \ \ \ \ \ \   v' {v'}^* = \half m^2\cos\galpha \ , 
\eeq
with the result 
\beq
E = \frac{1}{4} \tl\, L\, m^2\cos^2\galpha = 
\frac{\lambda m^2 }{4L}  \left(1-\frac{J^2}{L^2}\right) \ , 
\eeq
which is in agreement with \ci{Min2} (see also eq.(\ref{E(J,L)})). 
It is also instructive to substitute 
 this solution into  (\ref{LLL2}) and obtain the next
correction to the energy  which can be seen to agree with the expansion of the exact 
energy expression in 
(\ref{E(J,L)}).

Interpreted on  the spin chain side using  (\ref{opa}), the
 solution \rf{vsol} for  $\vv=v_m X^m $ 
thus determines the structure of the coherent SYM 
operator 
which is dual to the pulsating and rotating 
string solution.

\commentout{I take this out for the moment so paper looks complete
A more general special case is  of \rf{trunc} is when  $m$ runs 1,2,3,4, 
i.e. the string motion is within $S^3$ part of $S^5$. 
This model will contain both he 2-spin $SU(2)$ sector and the above $SO(3)$ 
model as two different special cases. 
Here we need to choose a useful parametrization of the 
 two unit 4-vectors $a_k$ and $b_k$ which are orthogonal to each other. 
We may fix a gauge as $b_4=0$; then $a_k$  can be parametrized by 3 angles 
of $S^3$ and a unit 3-vector $b_i$ ($i=1,2,3$) -- by  2 angles of $S^2$, 
with one of two angles  fixed by the orthogonality condition. 
An alternative parametrization corresponding to the gauge 
 $b_4=0$ gauge 
as follows:
...  so far action looks complicated...
$$ a_1= \sin \psi ( - \cos \p \cos \t \cos \a - \sin \p \sin \a) \ ,
\ \ \ \  \a_2 
= \sin \psi ( - \sin \p \cos \t \cos \a +  \cos \p \sin \a) \ ,$$
$$ a_3= \sin\psi  \sin\t \cos \a \ , \ \ \ \   a_4 =\cos \psi 
$$
$$ b_1= \cos \p \sin \t \ , \ \ b_2= \sin \p \sin \t \ , \ \ 
b_3= \cos \t , \ \  b_4=0$$ 
}


\section{Exact solution for pulsating and rotating string in $SO(4)$ sector \la{puls}
}

In this section we shall study the full classical 
 solution corresponding to a string rotating and pulsating in
the $S^3 $ part of $ S^5$. 
This solution was discussed in \cite{Min2}. 
Nevertheless,  we shall 
describe it here using a different (and straightforward) approach 
which appears to have  some 
technical and conceptual advantages. 
One important  point is that 
the use of action--angle variables will 
 allow us to obtain closed expressions 
for the relation between the energy and the conserved momenta.
 These expressions 
can be easily expanded at 
large energy to obtain the energy as a power series  
in the momenta and winding number, 
much in the same way as was done for the case of the two-spin 
rotating string in \cite{ft2,ft4,bfst}.

 Furthermore, we will 
 see that, in the limit of large energy, 
the rotation is along a maximal circle 
precessing around a fixed axis, which turns out to be precisely the motion  
described by the reduced sigma model of the previous section.

Before starting let us point out that the solution we are going to discuss 
is a special case of a more general class of pulsating and rotating solutions 
where the string may be stretched and rotating in all the 
three planes of $S^5$ \ci{art,tse2}. The corresponding 
integrable Neumann model
 is ``2-d dual'' ($\tau \leftrightarrow \sigma$) to the rotating string model 
\ci{art,tse2}. 
Let us choose  $t=\k \tau$ together with the following ansatz 
($i=1,2,3$)  \ci{art} 
\be\la{tat}
X_i = z_i(\tau)\ e^{i m_i \s} \ , \ \ \ \ \ \  z_i=  
r_i(\tau)\  e^{ i \a_i (\tau)} 
  \ , \ \ \ \ \ \ \ \ \sum^3_{i=1}  r^2_i(\tau) =1 \ ,  \ee
 where  the ``winding numbers''
 $m_i$ must take integer values
 in order  to satisfy the closed string periodicity 
 condition. 
  This ansatz  describes a pulsating and rotating string in $S^5$ 
  (special cases 
   were discussed previously in \ci{dev,gkp,Min1,larse,Min2}).
    The corresponding 
1-d effective Lagrangian is 
 \be
\label{Le}
\LL=\frac{1}{2}\sum^3_{i=1} ( \dot z_i \dot z^*_i -  m_i^2  z_i
z^*_i) +  \frac{1}{2} \Lambda(\sum^3_{i=1}  z_i z^*_i-1) \, .
\ee
Solving for $\dot \a_i$  we get  
$r^2_i \dot \a_i = \J_i$=const, where $\J_i$  are 
 the angular momenta.
 Then  we end up with   \be
\label{Lae}
\LL=\frac{1}{2}\sum^3_{i=1} \big(  \dot r^2_i   -   m_i^2  r^2_i -
   { \J_i^2 \ov r_i^2}
\big)  +   \frac{1}{2} \Lambda(\sum^3_{i=1}  r^2_i-1) \,  . 
\ee
Thus pulsating solutions carrying  3 spins $\J_i$ 
are   described by  a special 
Neumann-Rosochatius integrable system \ci{art}. 
Since  the  corresponding conformal
gauge constraints are  also $\tau  \leftrightarrow \s$ symmetric,  
they take a form similar  to the one in pure-rotation case
\be \la{cv}
\kappa^2=\sum_{i=1}^3( \dot r^2_i  + 
m_i^2 r^2_i  +  { \J_i^2 \ov r^2_i} )    \ , \ \ \ \ \ \ \ \ 
 \sum_{i=1}^3  m_i \J_i  =0  \ . \ee 
 One is thus to  look for   periodic  solutions of \rf{Lae} 
 subject to \rf{cv}, i.e. having finite 
 1-d energy. 
The general real 6-d  Neumann system  has the following 
six  commuting  integrals of motion 
($z_i=x_i + i x_{i+3}$):
\bea \la{inti} 
F_m=x_m^2+\sum^6_{m\neq n}\frac{(x_m x'_n-x_n x'_m)^2}
{w_m^2-w_n^2} \,  , \ \ \ \ \ \ \ \ \ \ 
\sum^6_{m=1}  F_m=1 \ , 
\eea
but  in the present case when  3 of the 6 frequencies are equal 
($w_i=w_{i+3}=m_i$) 
one needs to consider the  3 non-singular combinations of $F_m$ 
which then give the 3 integrals of \rf{Le}:
$
I_i = F_i + F_{i+3} ,$ or,  explicitly, 
 \be\la{nti}
I_i =r^2_i + \sum^3_{j\not=i} { 1 \ov m^2_i - m^2_j} 
\bigg[ (r_i r'_j - r_j r'_i)^2  + { \J^2_i \ov r^2_i} r^2_j 
+ { \J^2_j \ov r^2_j} r^2_i  \bigg]  \ . 
\ee
This gives two  additional (besides $\J_i$)  independent  integrals of
motion, explaining why this system is completely integrable.


\subsection{Form of the  solution}

In the simplest ``elliptic'' 
special  case we shall consider here 
the string is stretched in 34 plane, while rotating in 12 plane, i.e.
$m_2=m, \ m_1=m_3=0$ and $\J_1=\J, \ \J_2=\J_3=0$. This choice 
corresponds to a particle on $S^2$  with angular momentum 
$\J$ and an oscillator potential in the third direction. 
Equivalently, in terms of 6 real  coordinates this 
corresponds to the following ansatz:
$X_1= x_1(\tau), \ X_2=x_2(\tau), \ X_3+ i X_4= x_3(\tau) e^{i m \s}, 
\ X_{5,6}=0$, where $x_1= \sin \theta \ \cos \phi_1, \ x_2 = \sin \theta \ \sin \phi_1, 
\ x_3= \cos \theta$.
Explicitly, the 
  conformal gauge action for the string moving 
in $R_t \times S^3$ with the $S^3$ metric given 
in the angular parametrization by \rf{sss} is 
\be 
I= - \ha \sqrt \l \int d \tau \inti \left[
- (\del_p t)^2  + 
 (\del_p \theta) ^2 + \sin^2\theta\, (\del_p \phi_1)^2 + 
 \cos^2\theta\, (\del_p \phi_2)^2 \right] \ , 
\ee
and the resulting  equations of motion are satisfied by the ansatz 
\beq
t=\kappa\tau\ ,\ \ \ \theta=\theta(\tau)\ ,\ \ \ \phi_1=\phi_1(\tau)\ 
,\ \ \ \phi_2=m\sigma \ , 
\eeq
provided  the functions $\theta(\tau)$ and $\phi_1(\tau)$ satisfy
\beqa\la{ku}
\partial_\tau \left(\sin^2\theta\  \dot{\phi_1}\right) &=& 0 \ ,\\
\ddot{\theta}+(m^2+{\dot\phi_1}^2)\cos\theta\sin\theta &=& 0
\la{yu} \ . 
\eeqa 
These equations are equivalent to the conservation of the angular momentum
$J$ and the energy $E$
\be 
J=\sql\  P_{\phi_1} = \sql \J \ , \ \ \ \ \ \ \ \ \ \ \ \ 
E = \sql\ P_t = 
\sqrt{\lambda}\ \kappa
\ . 
\ee
The first equation implies
\beq
\dot{\phi}_1 = \frac{\J}{\sin^2\theta} \ . 
\eeq
The second can then be integrated once becoming 
 the conformal gauge constraint 
\beq
\dot{\theta}^2 + \frac{\J^2}{\sin^2\theta} + m^2 \cos^2\theta = \kappa^2 \ .
\label{eom1}
\eeq 
As already mentioned above,  this 
 equation can be interpreted as  describing  a periodic 
 motion of a particle 
of unit mass on  $S^2$ under the influence of 
an oscillator 
potential ${\cal U} (\theta)
 = \half m^2 x^2_3 = \ha m^2  \cos^2\theta$.

 Eq. (\ref{eom1}) can be integrated to obtain $\theta(t)$ as
\beq
\cos\theta = -a_-\ \mathbf{sn}(\frac{ma_+}{\kappa}t) \ ,
\label{thetasol}
\eeq
where $\mathbf{sn}$ denotes one of Jacobi's elliptic functions 
with modulus $k=a_-/a_+$, where  $a_{\pm}$ are 
constants of the motion given by
\beq\la{apm}
a^2_{\pm} = \frac{\kappa^2+m^2\pm\sqrt{(\kappa^2+m^2)^2-4m^2(\kappa^2-\J^2)}}{2m^2}
\ .\eeq
Notice that $a_{\pm}$ are real since $\kappa^2\ge \J^2$ as
 can be seen from  (\ref{eom1})
by considering the special point $\theta=\pi/2$. 
The motion of $\phi_1$ can be obtained by integrating 
$d\phi_1/d\theta$, which gives 
\beq
\phi_1 = -\frac{\J}{m a_+} \mathbf{\Pi}
\left(\arcsin(\frac{1}{a_-}\cos\theta(t)),a_-^2,\frac{a_-}{a_+}\right) \ , 
\label{phisol}
\eeq
where $\mathbf{\Pi}$ is a standard elliptic 
integral:\footnote{We follow the notation of \ci{Grad}.}
\beq
\mathbf{\Pi}(\psi,n,k) = \int_0^{\psi} \frac{d\b}{(1-n\sin^2\b)\sqrt{1-k^2\sin^2\b}} \ .
\eeq
 As we can see, the motion on the sphere is separable in 
the variables $\theta,\phi_1$. The system has two independent 
conserved momenta (being a special case of the Neumann
system, see above):  one is $J$ (which is the same as 
 the  action variable for $\phi_1$, \ 
$ I_{\phi_1}= \sql \int \frac{d\phi_1}{2\pi} P_{\phi_1} =J $), 
and the other
one  we can choose to be 
 the action variable corresponding to $\theta$:
\beqa
I_\theta &=& \sql  \oint \frac{d\theta}{2\pi} P_\theta = \sqrt{\lambda} \oint \frac{d\theta}{2\pi} 
\sqrt{\kappa^2-\frac{\J^2}{\sin^2\theta}-m^2\cos^2\theta} \la{Itheta} \\
  &=& \frac{4\sqrt{\lambda}}{2\pi}\frac{m}{a_+} 
\left[(a_-^2-1)\mathbf{K}\left(\frac{a_-}{a_+}\right)
+a_+^2\mathbf{E}\left(\frac{a_-}{a_+}\right)
+(a_+^2-1)(a_-^2-1)\mathbf{\Pi}\left(\frac{\pi}{2},a_-^2,\frac{a_-}{a_+}\right)\right] \ , \nonumber
\eeqa
where  $\mathbf{K}$, $\mathbf{E}$ and $\mathbf{\Pi}$
 are again the  standard elliptic integrals and the integration is over 
a complete period of motion which
takes places between a minimum and a maximum values of $\theta$. 
These  can be found  from the  zeros 
of the square root under  the integral.
 The functions $a_{\pm}$ are  defined in eq.(\ref{apm}).

 An interesting special 
case is when $J=0$ which is the pulsating string of \ci{Min1}. 
For $J=0$ eq. 
(\ref{apm}) implies 
 $a_+=\kappa/m$ and $a_-=1$. Then  (\ref{Itheta}) simplifies to
\beq\la{pulsr}
 \frac{L}{E} = \frac{I_{\theta}}{E} = \frac{2}{\pi}\, 
\mathbf{E}\left(\frac{\sql m}{E}\right),
\eeq
where we  used that $E= \sql \kappa$ and identified $L$ with $I_{\theta}$
 since $I_{\theta}$ is the only remaining conserved 
quantity.

\subsection{Expansion of the energy}

To compare with string theory we are interested in the limit of large energy, 
namely large $\kappa={E\ov \sql}$. 
The  expression \rf{Itheta} 
for the action variable $I_{\theta}$ can be expanded in this limit with the result
\beq
I_{\theta} = E-J -
 \frac{\lambda m^2}{4E} \left(1-\frac{J^2}{E^2}\right) + \ldots \ .
\eeq
 It is natural to introduce another 
conserved quantity which is the  sum of $I_{\theta}$ and $J$
\be  L\equiv I_{\theta}+J \ . \ee
To leading order of the large energy  expansion 
$L\equiv \sql \ell $ is equal to the  energy and thus
 $\ell\to \infty $  can 
be identified with an  analogous  expansion  parameter in \cite{Min2}. 
Inverting this expansion,  we obtain $E(J,L,\l)$ as
\beqa
E = L &+& \frac{\lambda m^2}{4\,L} \left(1-{J^2\ov L^2}\right) \bigg[
1  -  \frac{\lambda m^2}{16\,L^2}
\left(1+3{J^2\ov L^2}\right) + 
  \frac{\lambda^2 m^4}{64\,L^4}\
  \left(1-3{J^2\ov L^2}\right) \left(1-5{J^2\ov L^2}\right)  \nonumber \\ 
&-& \frac{\lambda^3 m^6}{4096\,L^6} 
 \left(13+113{J^2\ov L^2}-1017{J^4\ov L^4}+1211{J^6\ov L^6}\right)
 + \ldots \bigg]
\ .
\label{E(J,L)}
\eeqa
Here  we  computed several higher-order  terms in the expansion 
which  can be easily done 
with a computer algebra program.\footnote{While this paper
 was being written there appeared ref. \cite{Min3}  which also computed
 the first three leading terms of 
the  expansion of $E/L$ using a different 
 (and apparently more complicated) 
method. 
%
Our results agree.}
The energy or, more precisely, the ``energy density''
 $E/L$ 
 thus has an analytic expansion in $\tl = {\l\ov L^2}$ and $J\ov L$. 
The special case of $L=J$  when $E=J$ corresponds to a degenerate BPS 
point when the 
solution becomes a point-like geodesic: in this  case $I_\theta=0, \ \dot \theta
=0$, and then \rf{yu} implies that $\theta={\pi\ov 2}$, 
in which case the length of the string vanishes. 
Another case of interest is $J=0$ which
is the purely pulsating string of \ci{Min1}.

\subsection{Relation to leading-order solution}

It is instructive to analyze  the motion of the string in the large energy  limit. 
On  the sphere $(\theta,\phi_1)$ the string is seen as a point-like particle moving 
almost at the speed of light along a maximal  circle. The  $m^2 \cos^2 \theta$ 
potential produces an attractive force towards the equator. In each cycle the 
average force is zero but there is a torque which makes the averaged angular momentum precess around
 the $z$ axis as depicted in figure \ref{fig:precession}. This precession is described by the leading 
order solution.
\FIGURE{
\centerline{\epsfxsize=6cm\epsfbox{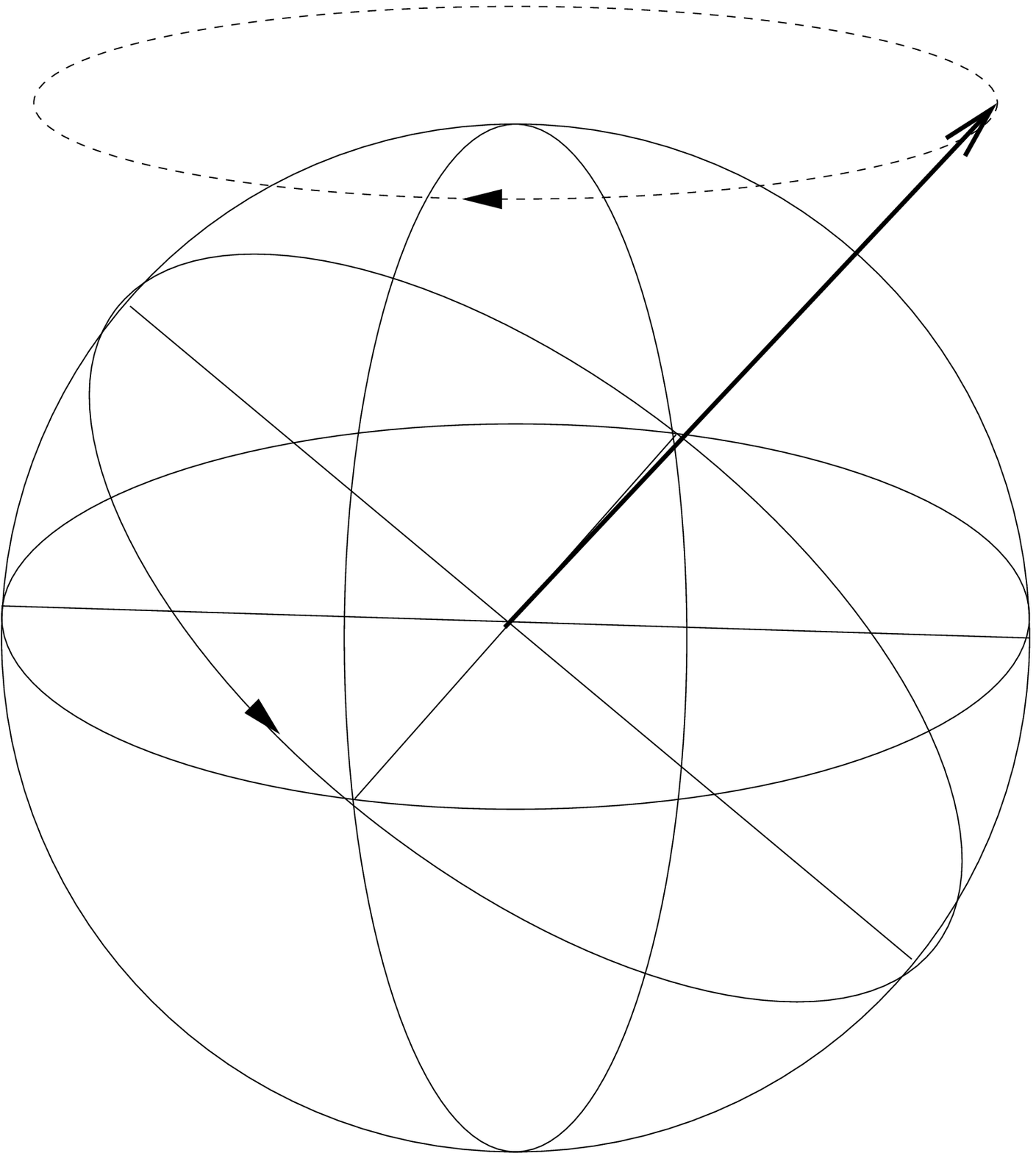}}
\caption{\small For large angular momentum the effective particle moves on the two-sphere parameterized 
by $(\theta,\phi_1)$ following a circle almost at the speed of light. The potential produces
 a force towards 
the equator which on average translates into a torque perpendicular to the angular momentum,
 causing a slow 
precession of the plane of rotation.}
\label{fig:precession}
}
The parameters characterizing the motion are the period of the motion and the rate of precession. 
The period $T$, or,  equivalently, the angular frequency $w=2\pi/T$ can be computed from eq.(\ref{thetasol})
using the fact that the elliptic function $\mathbf{sn}(u)$ with modulus $k$ is periodic 
with period $2\mathbf{K}(k)$. The result is:
\beq
\omega = \frac{2\pi}{T} = \frac{\pi ma_+}{2\kappa\mathbf{K}\left(\frac{a_-}{a_+}\right)}  
    \simeq 1 -   \frac{\lambda m^2}{4L^2}\left(1-3\frac{J^2}{L^2}\right)
    + \frac{\lambda^2m^4}{64L^4}\left(3+10\frac{J^2}{L^2}-21\frac{J^4}{L^4}\right) + \ldots,
\label{period}
\eeq
where we expanded in  large energy $\kappa$ and used eq.(\ref{E(J,L)}) 
to write the result as a function of
$L$ and $J$. Alternatively, we can use the fact
 that we have already expressed the energy as a function
of the action variables to compute the frequency as
\beq
\omega = \frac{\partial E}{\partial I_\theta} = \frac{\partial E}{\partial L} \ .
\eeq
Using (\ref{E(J,L)}) we then  reobtain (\ref{period}) in a simpler way.

The rate of precession can be found  by expanding (\ref{phisol}) for large $\kappa$ and computing the variation 
of $\phi_1$ in one cycle of $\theta$:  
\beqa
\Delta\phi_1 = \J \frac{4}{ma_+} \mathbf{\Pi}
\left(\frac{\pi}{2},a^2_-,\frac{a_-}{a_+}\right) 
              \simeq 2\pi \left[ 1-  \frac{\lambda m^2 J}{2L^3} 
              -  \frac{3\lambda^2 m^4 J}{16L^5}
 \left(1-3\frac{J^2}{L^2}\right) + \ldots \right] 
\eeqa
We see that there is a precession,  since, in one cycle, $\phi_1$ advances slightly 
less than $2\pi$. 
The rate of precession is therefore
\beq
\frac{1}{2\pi}\left(\Delta\phi_1-2\pi\right) = -\lambda  \frac{m^2 J}{2L^3} 
                   - \frac{3\lambda^2m^4J}{16L^5} \left(1-3\frac{J^2}{L^2}\right) + \ldots 
\label{precession}
\eeq
The sign here is in agreement with the direction of the torque
 (see fig. \ref{fig:precession}).
The above expression 
 can also be computed directly from eq.(\ref{E(J,L)}) 
using\footnote{This relation  is valid even if $\phi_1$ is not an angle variable, 
see next subsection for a more precise derivation.}
\beq
\Delta \phi_1 = \omega_{\phi_1} T = 2\pi \frac{\omega_{\phi_1}}{\omega} 
     = 2\pi \frac{\left(\frac{\partial E}{\partial J}\right)_{I_\theta}}{\left(\frac{\partial E}{\partial L}\right)_J}
     = 2\pi \left[ 1 + \frac{\left(\frac{\partial E}{\partial J}\right)_{L}}{\left(\frac{\partial E}{\partial L}\right)_J}
            \right] \ .
\eeq
Using again (\ref{E(J,L)}) we reproduce (\ref{precession}). 

We conclude  that, thanks to the 
identification of $J$ and $L$ with  the action variables, 
the function $E(J,L)$ 
can be  used to directly compute all the characteristic frequencies of the motion.

At leading order, these frequencies can be also 
obtained from the reduced sigma model \rf{joj} 
and,  therefore,
on the spin chain side, describe  the evolution  of the 
corresponding coherent state.

Having computed the relevant frequencies we can expand (\ref{thetasol}) at  large
 energy or small $\tilde \l$  to find that 
\beq
\cos\theta \simeq -\sqrt{1-\frac{J^2}{L^2}}\left\{\sin \omega t
+ \frac{\lambda m^2}{16 L^2}\left[\left(1-5\frac{J^2}{L^2}\right)\sin\omega t 
                                   + \left(1-\frac{J^2}{L^2}\right)\sin 3\omega t\right] +\ldots \right\} \ .
\label{thetaexp}
\eeq 
We can see here the typical properties of a perturbative expansion in classical mechanics. Higher
harmonics of the fundamental frequency $\omega$ appear in the higher order corrections. 
At the same
time,  the fundamental frequency itself has a perturbative 
expansion which  we obtained in eq.(\ref{period}).
Notice that attempting to  expand  (\ref{thetaexp}) 
in powers of $\tilde \lambda = {\lambda \ov L^2}$ 
by first plugging in  (\ref{period}) would
result in an  incorrect conclusion that there are perturbative terms linearly growing 
 in time. These are called
secular or resonant terms and should be eliminated, 
as was done here by correcting the fundamental frequency.
In the canonical formalism this is achieved by adding the average value (over a cycle) of the perturbative
Hamiltonian. This gives the corrected form of $E(J,L)$ which, as we saw, determines the corrected frequencies. 
 Furthermore, the angle-dependent part of the Hamiltonian
 is then eliminated by means of a canonical 
transformation. If we want to describe the system in terms of 
the original variables, undoing this canonical 
transformation produces 
 terms with higher harmonics of the fundamental frequency.

\subsection{Canonical perturbation \la{canpul}}

In the previous subsection we solved exactly the equation of motion for the pulsating solutions and
expanded them for large 
$\ell = {L\ov \sql}$ (with $J\ov L$ fixed). 
If one is interested just in the large $\ell $ expansion
it is possible to avoid the computation of the exact solution. 
Assuming that the only
$\sigma$ dependence is in $\phi_2(\sigma) = m\sigma$, and using conservation of $J$ to eliminate the 
variable $\phi_1$, we  reduce the problem  to the  study of  the Hamiltonian
\beq
 H = \half P_\theta^2 + \frac{1}{2} \frac{\J^2}{\sin^2\theta} +\half m^2\cos^2\theta \ .
\eeq
 For large $\ell$, the (bounded) 
 term proportional to $m^2$ can be consider as a perturbation and we can 
use the methods described in Appendix,  which, for the first order corrections, 
are actually simpler 
than the expansion of the previous subsection.
At lowest order, i.e. ignoring the term proportional to $m^2$, energy conservation
implies 
\beq
P_\theta = \sqrt{2H_0-\frac{\J^2}{\sin^2\theta}} \ ,
\eeq
where $H_0$ is the value  of $H$ at $m=0$. 
 This determines the action variable to be
\beq
I_\theta = \sql \oint {d \theta \ov 2 \pi} \sqrt{2H_0 -\frac{\J^2}{\sin^2\theta}} 
 = \sqrt{2\l H_0}-J \ .
\eeq
The canonical transformation from $(\theta,P_\theta)$ to action--angle variables 
$(\bar{\theta},I_\theta)$ is generated by a function $S(\theta,I_\theta)$ such that
$\bar{\theta}=\frac{\partial S}{\partial I_\theta}$, and $P_\theta=\frac{\partial S}{\partial\theta}$. Integrating
the latter  equation and then differentiating 
 the result with respect to $I_\theta$ we get
\beq
\bar{\theta} = \frac{\partial S}{\partial I_\theta} = 
  \int_{\frac{\pi}{2}}^{\theta} d \theta \frac{I_\theta+J}
{\sqrt{(I_\theta+J)^2-\frac{J^2}{\sin^2\theta}}}
     = \arcsin\left(\frac{L\cos\theta}{\sqrt{L^2-J^2}}\right) \ , \la{las}
\eeq
where we introduced 
$$L=I_\theta+J =\sql\   \ell \ ,\ \ \ \ \ \ \ \ell = \sqrt{2 H_0} \ . $$
 Equivalently,  we can write \rf{las} as 
\beq
\cos\theta = \sqrt{1-\frac{J^2}{L^2}}\sin\bar{\theta} \ .
\label{thetabar}
\eeq
Therefore,  the Hamiltonian becomes
\beq
H = \half \ell^2 + \frac{1}{4} m^2 \left(1-\frac{J^2}{L^2}\right) 
 -\frac{1}{4} m^2 \left(1-\frac{J^2}{L^2}\right) \cos2\bar{\theta} \ .
\eeq
The small parameter is 
${m\ov \ell} = { m \sql \ov L}$. The canonical method  reviewed in
 Appendix proceeds by 
choosing  a canonical transformation so that  to eliminate all $\bar{\theta}$ dependence
 in the transformed
Hamiltonian.  It is convenient to separate the perturbation $H_1$ into average 
$\langle H_1\rangle$ 
and fluctuating $\{H_1\}$ parts as follows:
\beqa
H_1 &=& \langle H_1\rangle + \{H_1\} =   \frac{1}{4} m^2 \left(1-\frac{J^2}{L^2}\right) 
 -\frac{1}{4} m^2 \left(1-\frac{J^2}{L^2}\right) \cos2\bar{\theta} \ , \\
\langle H_1 \rangle &=& \frac{1}{4} m^2 \left(1-\frac{J^2}{L^2}\right) \ ,\\
\{H_1\} &=&  -\frac{1}{4} m^2 \left(1-\frac{J^2}{L^2}\right) \cos2\bar{\theta} \ .
\eeqa
To second order the transformed Hamiltonian is
\beq
\bar{H} = H_0 + \langle H_1 \rangle + \half \langle\ [W_1,\{H_1\}]\ \rangle \ ,
\eeq
where $W_1$ is obtained from
\beq
[W_1,H_0] = \frac{\partial H_0}{\partial J}\frac{\partial W_1}{\partial\theta}= - \{H_1\} \ .
\eeq
This gives
\beq
W_1 = \frac{m^2}{8\ell}\left(1-\frac{J^2}{L^2}\right) \sin2\bar{\theta} \ .
\label{W1}
\eeq
For $H$ we obtain
\beq
H = \half {L^2\ov \l} + \frac{1}{4}\left(1-\frac{J^2}{L^2}\right) 
         - \frac{m^4}{64}\frac{\l^2}{L^4}\left(1-\frac{J^2}{L^2}\right)(5J^2-L^2) \ ,
\eeq
which is 
in agreement with eq.(\ref{E(J,L)}) as can be seen by noting that 
 $H=\half\kappa^2=\half {E^2\ov \l}$.

We can use the canonical transformation to obtain $\theta(\tau)$ as in (\ref{thetaexp}).
The first transformation is (\ref{thetabar}). 
Then we have to  use the transformation 
generated by $W_1$. At first order this gives
\beqa 
\cos\theta &\simeq& \sqrt{1-\frac{J^2}{L^2}}\sin\bar{\theta} + [W_1,\sqrt{1-\frac{J^2}{L^2}}\sin\bar{\theta} ] \non\\
  &=& \sqrt{1-\frac{J^2}{L^2}} \left\{ \sin\bar{\theta} 
      + \frac{\l m^2}{16L^2}\left[\left(1-5\frac{J^2}{L^2}\right)\sin\bar{\theta}
      +\left(1-\frac{J^2}{L^2}\right)\sin3\bar{\theta}\right]\right\} \ ,
\label{thetapert}
\eeqa 
where we used the value of $W_1$ given by (\ref{W1}). 
If we remember that $\bar{\theta} = \frac{\partial E}{\partial L} t$
then we see that we get precisely (\ref{thetaexp}).

We can also use this method to compute the time dependence of the angle $\phi_1$. 
The first point  to notice
is that although $J$ is an action variable, $\phi_1$ is not the corresponding angle variable as can easily
be seen since $\dot{\phi_1}$ is not constant. The angle variable can be obtained by solving the Hamilton-Jacobi 
equation which gives the action
\beq
S = J \phi_1 + \int d\theta \sqrt{L^2-\frac{J^2}{\sin^2\theta}} \ ,
\eeq
which was already discussed. We just added the $J\phi_1$ term. The angle variable can be obtained now
as
\beq
\bar{\phi}_1 = \frac{\partial S}{\partial J} = \phi_1 + \bar{\theta} - \arctan\left(\frac{J}{L}\tan\bar{\theta}\right) \ ,
\eeq
where we used  that $L$ is also a function of $J$ through $L=I_\theta+J$.  This gives rise to the
term $\bar{\theta} =\frac{\partial S}{\partial L}$. Inverting this we get 
the expression for  $\phi_1$ in 
terms of action--angle variables:
\beq
\phi_1 = \bar{\varphi} + \arctan\left(\frac{J}{L}\tan\bar{\theta}\right) \ ,
\eeq
where we defined $\bar{\varphi} = \bar{\phi_1} -\bar{\theta}$ which is the 
variable conjugate to $J$ at constant  $L$.
Now we can very easily compute the time dependence of $\phi_1$ at first order:
\beq
\phi_1 = \omega_{\bar{\varphi}} t  + \arctan\left(\frac{J}{L}\tan\omega t\right) 
           + [W_1,\bar{\varphi} + \arctan\left(\frac{J}{L}\tan\bar{\theta}\right)] \ ,
\eeq
 where we introduced the frequencies $\omega_{\bar{\varphi}}= 
\left(\frac{\partial E}{\partial J}\right)_L $ and 
$\omega=\frac{\partial E}{\partial L}$ determining the time evolution of the
angle variables $\bar{\varphi}=\omega_{\bar{\varphi}} t$  and  $\bar{\theta}=\omega t$. 
Evaluating the Poisson bracket in the last term we obtain the final result:
\beq
\phi_1 \simeq \omega_{\bar{\varphi}} t + \arctan\left(\frac{J}{L}\tan \omega t \right) 
   + \frac{\l m^2}{4L^2}\frac{J}{L}\left(1-\frac{J^2}{L^2}\right)
       \frac{\sin2\omega t}{1-\left(1-\frac{J^2}{L^2}\right)\sin^2\omega t} \ .
\label{phi1pert}
\eeq
It is instructive and useful for our purpose to
 put together eqs.(\ref{thetapert}) and (\ref{phi1pert}) by 
using the parameterization
\beqa
X_1 &=& \sin\theta \cos\phi_1 , \ \ X_2 = \sin\theta \sin\phi_1\ , \non \\
X_3 &=& \cos\theta \cos\phi_2 ,\ \ X_4 = \cos\theta \sin\phi_2 \ .
\eeqa
To obtain the time dependence of these coordinates we first use that
\beqa
\cos\left[\arctan\left(\frac{J}{L}\tan\omega t\right) \right] &=& 
   \frac{\cos\omega t}{\sqrt{1-\left(1-\frac{J^2}{L^2}\right)\sin^2\omega t}} \ ,  \\
\sin\left[\arctan\left(\frac{J}{L}\tan\omega t\right) \right] &=& 
  \frac{J}{L} \frac{\sin\omega t}{\sqrt{1-\left(1-\frac{J^2}{L^2}\right)\sin^2\omega t}} \ ,
\eeqa
to expand $\cos\phi_1$ and $\sin\phi_1$. Also, from (\ref{thetapert}) we get that 
\beq
\sin\theta \simeq  \sqrt{1-\left(1-\frac{J^2}{L^2}\right)\sin^2\omega t} + \cO\left(m^2\right) \ .
\eeq
As a result, 
\beqa
X_1 &=& \cos\omega_{\bar{\varphi}} t\  \cos\omega t - \frac{J}{L} \sin\omega_{\bar{\varphi}}t\  \sin\omega t \ ,\\
X_2 &=& \sin\omega_{\bar{\varphi}} t\  \cos\omega t + \frac{J}{L} \cos\omega_{\bar{\varphi}}t\ 
 \sin\omega t \ ,\\
X_3 &=& \sqrt{1-\frac{J^2}{L^2}} \sin\omega t\  \cos m\sigma \ ,\\
X_4 &=& \sqrt{1-\frac{J^2}{L^2}} \sin\omega t\  \sin m\sigma \ ,
\eeqa
where we used that $\phi_2 = m\sigma$ and dropped all terms of order $m^2$ except in the frequencies $\omega$ and
$\omega_{\bar{\varphi}}$, i.e.   in the secular terms. These contribution should be described by the 
averaged Hamiltonian. To compare with the results
 of section \ref{particularsolution} let us introduce
the notation $\sin\galpha={J\ov L}$ and then use 
the parameterization of eq.(\ref{geo}),  namely
\beq
X_m =  a_m \cos\omega t + b_m \sin\omega t  \ ,
\eeq
where we put $\alpha=\omega t$. We can then identify
\beqa
\vec{a} &=& (\cos\omega_{\bar{\varphi}} t,\ \sin\omega_{\bar{\varphi}} t,\ 0,\ 0) \ , \non\\
\vec{b} &=& (-\sin\galpha \sin\omega_{\bar{\varphi}} t,\ \sin\galpha \cos\omega_{\bar{\varphi}} 
t,\ \cos\galpha\cos m\sigma,
             \ \cos\galpha\sin m\sigma) \ ,
\eeqa
and thus get the following expression for  $V$ defined as in (\ref{fas}), i.e.
$V= \frac{1}{\sqrt{2}} (\vec{a}-i\vec{b})$:
$$
V 
= -\frac{i}{\sqrt{2}} (-\sin\galpha\sin\omega_{\bar{\varphi}}t+
i\cos\omega_{\bar{\varphi}} t, \  \sin\galpha\cos\omega_{\bar{\varphi}} t
     +i\sin\omega_{\bar{\varphi}} t,\ \cos\galpha\cos m\sigma,\ 
\cos \galpha \sin m\sigma) 
$$
This is in  perfect agreement with (\ref{vsol}) which was obtained 
by  solving the equations of the reduced sigma model. The only difference
is an irrelevant overall factor $i$.
 The frequencies also agree since from eq.(\ref{E(J,L)}) we get 
\beq
\omega_{\bar{\varphi}} = \left.\frac{\partial E}{\partial J}\right|_L \simeq -
 \frac{\l m^2}{2L^2}\frac{J}{L} \ .
\eeq
To compare with $\omega$ in  (\ref{ABw}) we should note that 
because of the rescaling of time by $\tilde \l$ we need 
to rescale the frequencies or  $m\rightarrow { \sqrt{\lambda}\ov L} m $. Also, we should use 
that  $\sin\galpha=\frac{J}{L}$.

\section{Conclusions \la{conclu}}

 In this paper we studied strings that move fast, close to the speed of light, 
in the $S^5$ part  of \adss
and compared them to semiclassical states of the $SO(6)$ spin chain 
Hamiltonian. The latter is  
 equivalent  \ci{mz1} to the 1-loop, large-$N$,
dilatation operator acting on the scalar sector of \N{4} SYM. 
We showed how the classical string action can be systematically 
expanded in this limit, first for rotating strings and then in general.
 The leading order result agrees with the sigma model that we derived
 from the field theory 
by  restricting  to operators satisfying a ``locally BPS'' condition which leads 
to a  Grassmanian $G_{2,6}$ sigma model \ci{ST,mikk}.

As a result,  we get a precise mapping between operators and strings \ci{bmn,kru,krt,mikk}. 
In the present  case, the operators are products of a large number of scalar fields. The map is such
that to each portion of the string   one associates a 
particular linear combination of the scalar fields in the product.
 This is done
by identifying $\sigma$, the coordinate along the string, with the discrete coordinate which 
labels the scalar fields. In the limit \rf{cont}
 the trajectory of each point of the string is approximately
a maximum circle characterized by an $SO(6)$ angular momentum which agrees
with the $SO(6)$ R-charge carried by the associated scalar field. Furthermore, the maximum circle along 
which each point of the string is moving changes in time, 
 and correspondingly the scalar operator also 
changes. This motion is the one described by the sigma model derived from the spin chain Hamiltonian 
or from the expansion of the string action. 

 Our  result generalizes  similar comparisons done for rotating \ci{bfst,kru,krt} and pulsating 
\ci{Min1,Min2}
strings, not only because it   provides a map between the dual 
configurations, but also 
because it includes many more examples, for instance, fluctuations around the known 
solutions.
 
 It seems feasible that we can extend this comparison to two loops. For the $SU(3)$ sector the
necessary ingredients are in \ci{beit} and in section \ref{SU(3)} of this paper.
In addition, one needs to use the important  observation of \ci{Min3}
that mixings with fermionic states can be ignored to 
leading order in $1/L$ expansion. 
For $SO(6)$, the two-loop dilatation
 operator needs to be computed.\foot{Again, the result 
of  \ci{Min3} 
suggest that if one is interested only in the large $L$ limit 
of this dilatation operator its  computation can be simplified 
since the operator mixing should be  suppressed for  $L\rightarrow\infty$.}
  Once the two-loop dilatation operator is known, 
the field theory calculation should  reduce
 to a computation of quantum corrections 
in the spin chain as in \ci{krt}.

 On the string side
the expansion to higher orders is classical and was developed here 
in great detail using the following method.
 First,  one isolates a fast variable $\alpha$ such that the velocities 
of all the other coordinates can be considered small.
 In order to do so we have to use  {\it phase space} description 
since the fast variable cannot be isolated in coordinate space.
 The basic reason is that the fast variable 
is the polar angle in the plane determined by the position {\it and} 
the  momentum (see also  \ci{mikk} 
for a similar approach).

 The momentum conjugate to $\alpha$, which we call $P_{\a}$, is large and can be used to expand the 
action as a power series in $1/P_\a$. However, since the (phase-space) Lagrangian depends on $\a$,
the momentum $P_\a$ is not conserved. To get a conserved quantity we do a series of canonical  
transformations that eliminate the dependence on $\alpha$ order by order in $1/P_\a$. 

 The new $P_{\a}$ is conserved and can be used to characterize the solutions. It can also be gauge 
fixed to a constant as can be seen more intuitively by doing a T-duality transformation along the 
fast variable and then fixing a static gauge. In this gauge $P_{\a}$ is uniformly
distributed along the string and therefore can be naturally associated with the length of
the spin chain. This is the parameter that is usually denoted as $J$ for the rotating strings and $L$ for the
pulsating ones.
 
 Expanding the final Lagrangian for large $P_\a$ is then a simple Taylor expansion.
We emphasize that the difficult step is the previous one, namely, to eliminate
 the fast variable. This we did here 
 only to the  next to the  leading order. As a technical point,  we
 found it  convenient to 
use the embedding Cartesian coordinates and also to
 introduce the fast variable $\alpha$ as a redundant 
coordinate (giving rise to a $U(1)$ gauge invariance).

 The whole procedure relies on using a gauge  and a set of variables which 
are  well adapted
to the expansion at the classical level. It would be interesting to study
if similar simplifications occur in quantum string 
 calculations like  the ones in \ci{tse1,ft3}. 

 As a  check,  we studied  several particular 
 examples, including pulsating and rotating 
solutions and were able to verify  the agreement with previously 
known   results \ci{bfst,kru,lopez,ST,Min1,Min2}. 
This also allowed  us to obtain the coherent operators that correspond to the pulsating solutions 
of \ci{Min1,Min2}. 
These are not equivalent to  the eigenstates of the dilatation operator 
that are related to the same  string  solutions 
(have the same energy) 
and which were identified 
in \ci{Min1,Min2} using the Bethe ansatz. 
We emphasize that the operators we discussed 
here are not eigenstates of the dilatation operator, rather, 
they represent semiclassical or coherent SYM states   which are naturally 
associated to  the semiclassical string states 
described  by the classical string solutions.

Let us mention also 
another result we obtained for the pulsating solutions:  the exact 
relation between the classical  energy and the conserved quantities $J$
and $L$. 
The solution and its energy  can be written in 
terms of elliptic integrals much in the same way as was previously done 
 for the 2-spin rotating string solutions.
 Similarly,  the large $L\ov \sql $ expansion becomes
 just a Taylor expansion and can be easily carried out. 
All 
this is achieved by identifying $J$ 
and $L$ with the action variables conjugated to the 
corresponding angles. 
This  identification  has also  the advantage of allowing the calculation
of the characteristic frequencies of  motion
 as derivatives of the energy with
respect to the action variables $J$ and $L$.

Finally, from a  broader perspective, the importance of understanding 
 the duality  correspondence 
for a large class of time-dependent semiclassical states 
is that  it may show   the way to the study of 
generic string oscillation modes and thus further clarify the structure 
of \adss string spectrum. One  lesson we have learned here is 
the crucial role  of the  phase space approach:  the same 
{\it local}  phase space effective sigma model action
in which coordinates and momenta appear on an equal footing 
 emerged from both  the spin chain and  the string theory.
Eliminating momenta would lead to a nonlocal action.

\acknowledgments
We are grateful to 
A. Lawrence, E. Lopez, 
D. Mateos,
A. Mikhailov, J. Minahan,   
N. Nekrasov,  B. Stefanski and K. Zarembo 
  for useful  suggestions,  discussions and comments.
The  work of M.K. was supported in 
part by NSF through grants PHY-0331516, PHY99-73935 and DOE 
under grant DE-FG02-92ER40706. The work of 
A.T. was supported  by the  DOE
grant DE-FG02-91ER40690, the INTAS contract 03-51-6346
and the RS Wolfson award.

\appendix

\section{Canonical perturbation theory \la{app}}

In this appendix we give a brief introduction to the subject of canonical perturbation theory.
The main purpose is to familiarize the reader with the notation used in the main text and to
make the paper self-contained. We follow closely \cite{LL} where the interested reader can find
many more details, examples and references to the original literature.
 The case of non-canonical
variables which is the one we studied in 
 the main text is discussed \eg\ in \ci{LJ}. 

 Since the equations of motion for a dynamical system cannot usually be integrated easily,
perturbation theory is a widely used and well studied method. 
Let us consider some cases in which it can be applied.
 Suppose first that we have an integrable system with canonical variables $\bar{x}=(p_i,q_i)$, Hamiltonian  
$H(p,q)$ and Poisson brackets
\beq
{[}f,g] = \frac{\partial f}{\partial q_i}\frac{\partial g }{\partial p_i}
   -\frac{\partial f}{\partial p_i}\frac{\partial g}{\partial q_i} \ . 
\eeq
 We can do a canonical transformation to angle-action
 variables $(\bar{\theta}_i,J_i)$ such that the
Hamiltonian becomes a function of the $J_i$ only: $H(J_i)$. Hamilton's equations $\frac{d\bar{x}}{dt} = [\bar{x},H]$
can now be easily solved:
\beq
J_i = \mbox{const}\ ,\ \  \ \ \ \bar{\theta}_i = \omega_i t + \beta_i\ , 
\ \ \ \ \ \mbox{with} \ \ 
 \omega_i=\frac{\partial H}{\partial J_i}  \ .
\eeq
Finding such a transformation,  however,  is not simple. If, instead, we are able to find a transformation 
$(p_i,q_i)\rightarrow (J_i,\theta_i)$ that puts the Hamiltonian in the form 
\beq
H=H_0(J_i)+\epsilon H_1(J_i,\theta_i) + \epsilon^2 H_2(J_i,\theta_i) + \ldots 
\eeq
with $\epsilon\ll 1$, then  we can use perturbation theory
in $\epsilon$ 
 to find a further transformation 
$(J_i,\theta_i)\rightarrow (\bar{J}_i,\bar{\theta}_i)$ such that the 
Hamiltonian becomes  a function of the
$\bar{J}_i$ only. The ultimate objectives are: 
(i)  to determine the characteristic frequencies 
 $\omega_i=\partial H(\bar{J}_i)/\partial \bar{J}_i$,
and (ii)   to use the canonical transformation
$p(\bar{\theta}_i,\bar{J}_i), q(\bar{\theta}_i,\bar{J}_i)$ to 
determine how these frequencies appear
in the expressions for the 
 original coordinates. All this can be done order by order in $\epsilon$.

Below we will show how this can be done for a one-dimensional system which is always integrable. 
Another related example is when the motion is such that one variable 
changes  much faster than 
the others. By considering the other variables  ``frozen'' we can 
reduce the system to a one-dimensional 
one which is integrable and can be again 
transformed to action--angle variables  perturbatively. This 
case is similar to the previous one only that now we should consider the motion of the other 
variables as perturbations. Notice that in this case it is not necessary, and can actually be inconvenient,
to use action--angle variables for the ``slow''  coordinates.

Let us also point out that while  most systems are not 
integrable,  perturbation theory can sometimes also be 
used to approximate them by an integrable one. Although, generally speaking, this is the most 
interesting case, it will not concern us here.
 
 Before going into further details let us discuss which is the first problem that
one faces when doing perturbation theory. 
The problem  is due to the so called secular terms and is 
solved by the canonical method.\footnote{This not the only
 problem that perturbation theory has to face. 
Another well-known problem that we will  not discuss
 here is related to  the so called small denominators.}
We can illustrate it with a simple example. 
Consider  the motion   which is described by the function
\beq
y(t) = a \sin \omega t \ ,
\eeq
but where  we know $a$ and $\omega$ only as 
 expansions $a=a_0 + \epsilon a_1+\ldots$, 
$\omega=\omega_0+\epsilon \omega_1 +\ldots$.
At leading  order the best we can do is to approximate
\beq
y(t) \simeq a_0 \sin(\omega_0 t+ \epsilon\omega_1t) + \epsilon a_1\sin(\omega_0 t+ \epsilon\omega_1t) \ .
\eeq
It would be incorrect to further expand in $\epsilon$ to get
\beq
y(t) \simeq a_0 \sin\omega_0 t  + \epsilon a_0 \omega_1 t\cos\omega_0 t +  \epsilon a_1\sin(\omega_0 t) \ .
\label{incorrect}
\eeq
This is clear since for large $t$ we cannot guarantee that $ \epsilon \omega_1 t$ would be small.
The result of the incorrect procedure is a perturbation linearly growing in $t$ which is usually
called a secular term, using nomenclature from celestial mechanics. This obvious point can become 
slightly more confusing if we now suppose that $y(t)$ is obtained by solving an equation of motion. The zero 
order solution would be $y_0(t) = a_0 \sin\omega_0 t $ and to first order  we would be tempted to use the ansatz
\beq
y(t) \simeq  a_0 \sin\omega_0 t + \epsilon y_1(t) \ .
\eeq
Solving for $y_1(t)$ we are bound to obtain
 (\ref{incorrect}), namely the incorrect result. The natural 
solution is to use the 
 ansatz where $\omega_0$ is replaced by $\omega_0+\epsilon \omega_1$ 
with $\omega_1$ chosen so as 
 to cancel the term linear in $t$. The physical interpretation of this is clear.
If we think of two particles, one following an
 unperturbed trajectory and the other the perturbed one,
even though the trajectories are close,
 one particle lags behind the other because of the frequency
correction, producing 
 the linear term in the difference of positions.

Canonical perturbation theory deals with this problem in the following way. 
At each order in perturbation theory
one has a Hamiltonian $H_n(\theta,J)$ which can be divided into 
an averaged and a  fluctuating part  as follows 
\beqa
 H_n(\theta,J) &=& \langle H_n \rangle (J) + \{ H_n\}(\theta,J) \ , \non\\
 \langle H_n \rangle (J) &=& \frac{1}{2\pi} \int_0^{2\pi} d\theta\   H_n(\theta,J) \ , \\
  \{ H_n\}(\theta,J) &=& H_n - \langle H_n \rangle (J) \ . \non
\la{avfl}
\eeqa
The fluctuating part can be eliminated by a canonical transformation to get a Hamiltonian depending
on $J$ only (at this order). If we try to eliminate also $\langle H_n \rangle (J)$ this gives
rise to secular terms. Instead, since $\langle H_n \rangle (J)$ depends only on $J$, it
can be incorporated into the transformed Hamiltonian and therefore gives a correction to the 
frequency $\omega=\frac{\partial H}{\partial J}$.

Let us now consider  this procedure in more detail. We have a system with canonical variables 
$(\theta,J)$ that we want to transform into $(\bar{\theta},\bar{J})$ such that the Hamiltonian
\beq
H=H_0 (J)+\epsilon H_1(J,\theta) + \epsilon^2 H_2(J,\theta) + \ldots \ 
\la{HJt}
\eeq
becomes independent of $\bar{\theta}$. Denoting $\bar{x}=(J,\theta)$, 
such  transformation $T_\epsilon:
\bar{x} \rightarrow \bar{x}_\epsilon$ will be generated by 
a function $W(\bar{x},\epsilon)$ 
\beq
\frac{d\bar{x}}{d\epsilon} = [\bar{x}, W(\bar{x},\epsilon)] \ .
\la{xeq}
\eeq
Solving for $\bar{x}(\epsilon)$ gives   a  
 canonical transformation $\bar{x}(0)\rightarrow \bar{x}(\epsilon)$. 
For small $\epsilon$ we can expand   $W$ as
\beq
W = W_1 + \epsilon W_2 +  \ldots \ .
\eeq
Expanding also $\bar{x}$ in powers of $\epsilon$ and taking into
 account that the functions $W_{n}$ should be further expanded
since they depend on $\bar{x}$, we can solve (\ref{xeq}) obtaining 
\beq
\bar{x}(\epsilon) = \bar{x} - \epsilon [W_1,\bar{x}] -\half \epsilon^2[W_2,\bar{x}] 
    +\half \epsilon^2 [W_1,[W_1,\bar{x}]] + \cO(\epsilon^3) \ ,
\la{xs}
\eeq 
where on the right hand side all functions are evaluated at $\epsilon=0$.
An arbitrary function $F(\bar{x})$ transforms in such a way that $F_{\epsilon}(\bar{x}_{\epsilon})=F(\bar{x})$.
This condition and (\ref{xs}) imply that
\beq
F_\epsilon = F + \epsilon [W_1,F] + \half \epsilon^2 [W_2,F] +\half \epsilon^2 [W_1,[W_1,F]] + \cO(\epsilon^3) \ .
\eeq
If we now put $F=H$ with $H$ being 
 the Hamiltonian of eq.(\ref{HJt}) we get the transformed Hamiltonian 
\beq
H_{\epsilon} = H_0 + \epsilon H_1 +\epsilon^2 H_2 
  + \epsilon [W_1,H_0] + \epsilon^2 [W_1,H_1] + \half \epsilon^2 [W_2,H_0] +\half \epsilon^2 [W_1[W_1,H_0]] 
  + \cO(\epsilon^3) \ .
\eeq
The left hand side should be independent of $\theta$ so we should choose $W_1$ and $W_2$ to cancel the
$\theta$ dependence on the right hand side. As discussed before, we separate each term on the right hand side
into averaged and fluctuating parts using (\ref{avfl}) in order to avoid secular terms.
This leads to the equations
\beqa
{[}W_1,H_0] &=& \frac{\partial H_0}{\partial J}  \frac{\partial W_1}{\partial\theta} = -\{H_1\} , \\ 
{[}W_2,H_0] &=& \frac{\partial H_0}{\partial J}  \frac{\partial W_2}{\partial\theta} = -2 \{H_2\} - 2[W_1,\langle H_1\rangle]
     -[W_1, \{H_1\}] , \ \ 
\la{Weq}
\eeqa
which determine $W_1$ and $W_2$. Notice that attempting to eliminate 
the average value $\langle H_1 \rangle$
give rise to a term linear in $\theta$ in $W_1$, namely a secular term (since $\theta=\omega t$). 
The transformed Hamiltonian is given by the averaged terms that remain
 after the transformation
\beq
\bar{H} = H_0 +\epsilon \langle H_1\rangle + \epsilon^2 \langle H_2\rangle + \half\langle[W_1,\{H_1\}]\rangle  \ ,
\la{Hpeq}
\eeq
and it determines the frequencies. Inverting the canonical transformation 
(\ref{xs})  
allows one  to immediately write down the time dependence of the original variables. 

 As a final comment,  let us point out that
 although, at first sight, this procedure may  seem involved,
it easy to convince oneself by working out
 an example that it is much easier than a more
straightforward approach. One of the main points is
 that the coordinate transformations used 
are canonical and therefore are defined 
 by just one scalar function. Trying to solve the equations directly
requires  obtaining a separate  function for each of the coordinates. 
This soon becomes intractable. 
 A  very simple example is actually the one we worked out 
in section \ref{canpul} and it can 
be used to illustrate all the points discussed here.


\end{document}